\documentclass[floats,floatfix,showpacs,amssymb,prd,nofootinbib,aps,notitlepage,reprint]{revtex4-1}
\pdfoutput=1
\usepackage{graphicx}
\usepackage[utf8]{inputenc}
\usepackage{graphicx, epsfig, amssymb} 
\usepackage{amsmath, amsfonts}
\usepackage{bm} 
\usepackage[normalem]{ulem} 
\usepackage{soul}
\usepackage[linktocpage]{hyperref}
\usepackage[caption=false]{subfig}
\usepackage[usenames]{color}
\usepackage{tikz}
\usetikzlibrary{arrows,shapes}
\usetikzlibrary{trees}
\usetikzlibrary{matrix,arrows} 				
\usetikzlibrary{positioning}				
\usetikzlibrary{calc,through}				
\usetikzlibrary{decorations.pathreplacing}  
\usepackage{pgffor}							

\usetikzlibrary{decorations.pathmorphing}	
\usetikzlibrary{decorations.markings}
\tikzset{
    vector/.style={decorate, decoration={snake}, draw},
	provector/.style={decorate, decoration={snake,amplitude=2.5pt}, draw},
	antivector/.style={decorate, decoration={snake,amplitude=-2.5pt}, draw},
    fermion/.style={draw=black, postaction={decorate},
        decoration={markings,mark=at position .55 with {\arrow[draw=black]{>}}}},
    fermionbar/.style={draw=black, postaction={decorate},
        decoration={markings,mark=at position .55 with {\arrow[draw=black]{<}}}},
    fermionnoarrow/.style={draw=black},
    gluon/.style={decorate, draw=black,
        decoration={coil,amplitude=4pt, segment length=5pt}},
    scalar/.style={dashed,draw=black, postaction={decorate},
        decoration={markings,mark=at position .55 with {\arrow[draw=black]{>}}}},
    scalarbar/.style={dashed,draw=black, postaction={decorate},
        decoration={markings,mark=at position .55 with {\arrow[draw=black]{<}}}},
        zeroscalar/.style={dotted,draw=black, postaction={decorate},
        decoration={markings,mark=at position .55 with {\arrow[draw=black]{>}}}},
    zeroscalarbar/.style={dotted,draw=black, postaction={decorate},
        decoration={markings,mark=at position .55 with {\arrow[draw=black]{<}}}},
    scalarnoarrow/.style={dashed,draw=black},
    electron/.style={draw=black, postaction={decorate},
        decoration={markings,mark=at position .55 with {\arrow[draw=black]{>}}}},
	bigvector/.style={decorate, decoration={snake,amplitude=4pt}, draw},
}

\tikzstyle{block} = [draw, rectangle,
    minimum height=3em, minimum width=6em]

\def\be{\begin{equation}}
\def\ee{\end{equation}}

\newcommand{\bea}{\begin{eqnarray}}
\newcommand{\eea}{\end{eqnarray}}
\newcommand{\ben}{\begin{enumerate}}
\newcommand{\een}{\end{enumerate}}
\newcommand{\bi}{\begin{itemize}}
\newcommand{\ei}{\end{itemize}}

\def\ga{\mathrel{\raise.3ex\hbox{$>$\kern-.75em\lower1ex\hbox{$\sim$}}}}
	\def\la{\mathrel{\raise.3ex\hbox{$<$\kern-.75em\lower1ex\hbox{$\sim$}}}}

\def\be{\begin{equation}}
\def\ee{\end{equation}}

\def\I_M{{I_{\scriptscriptstyle M\times M}}}

\def\be{\begin{equation}}
\def\ee{\end{equation}}
\def\bea{\begin{eqnarray}}
\def\eea{\end{eqnarray}}
\newcommand{\beq}{\begin{eqnarray}}
\newcommand{\eeq}{\end{eqnarray}}

\newcommand{\beqal}{\begin{eqnarray}\label}
\newcommand{\none}{\end{eqnarray}}
\newcommand{\beqa}{\begin{eqnarray}}
\newcommand{\eeqa}{\end{eqnarray}}

\begin{document}\setlength{\unitlength}{1mm}
\title{\large Removing the Faddeev-Popov zero modes from Yang-Mills theory in spacetimes with compact spatial sections}

\author{Jos Gibbons}\email{jg1047@york.ac.uk}
\author{Atsushi Higuchi}\email{atsushi.higuchi@york.ac.uk}
\affiliation{Department of Mathematics, University of York, Heslington, York, YO10 5DD, United Kingdom}

\begin{abstract}	
It is well known that in de~Sitter space the (free) minimally coupled massless scalar field theory does not admit any de~Sitter-invariant Hadamard state.
Related to this is the fact that
the propagator for the massive scalar field corresponding to the de~Sitter-invariant vacuum state diverges in the massless limit, with the
infrared-divergent term being a
constant.  Since the Faddeev-Popov ghosts for the covariantly quantized Yang-Mills theory are minimally coupled massless
scalar fields, it might appear that de~Sitter symmetry would be broken in the ghost sector of Yang-Mills theory in de~Sitter space.  It is shown in this paper that
the modes responsible for de~Sitter symmetry breaking can be removed in a way consistent with BRST invariance and that a de~Sitter-invariant theory can be
constructed.  More generally, it is shown that the spatially constant modes (the zero modes) of the Faddeev-Popov ghosts can be disposed of in a wide class of
spacetimes with compact spatial sections.  Then, the effective
theory obtained by removing the zero modes, which contains a nonlocal interaction term, is shown
to be equivalent to the theory corresponding to using a Faddeev-Popov-ghost
propagator with the constant infrared-divergent term removed, provided that one
can freely integrate by parts in the spacetime integral at the vertex for the ghost interaction term.
\end{abstract}

\pacs{04.62.+v, 11.10.Ef, 11.15.Bt, 03.70.+k}

\date{January 7, 2015}

\maketitle


\section{Introduction}

Physics in de~Sitter space has been studied extensively since inflationary cosmology was proposed in the early 1980s~\cite{inf1,inf2,inf3,inf4,inf5}
because this spacetime is a very good approximation to the geometry of the Inflationary Universe.
Recently the detection of remnants of the primordial gravitational waves was reported~\cite{BICEP2}.  If confirmed by further observations,
this will provide  strong evidence for an
inflationary phase of our Universe and, hence,
for physical relevance of de~Sitter space.  Moreover,  the expansion of our Universe is believed to be
accelerating~\cite{accel1,accel2}, and it may eventually expand exponentially, thus becoming approximately de~Sitter space.   The proposal of the dS/CFT
correspondence~\cite{strominger2001} gives another motivation to study physics in de~Sitter space.  Thus, it will be useful to understand properties of
Yang-Mills theory, which is an important ingredient of any realistic model of particle physics, in de~Sitter space.

A method commonly used to quantize Yang-Mills theory is the canonical quantization that starts from the Lagrangian density
with a covariant gauge-fixing term and the
corresponding Faddeev-Popov (FP) term~\cite{FP}.  In this method of quantization it can readily be seen that the FP ghosts, i.e.\ the FP ghost and
antighost, are minimally coupled massless scalar
fields.  Now, it is well known that the (free) minimally coupled massless scalar field theory suffers from infrared (IR)
divergences in de~Sitter space~\cite{FordParker1977a}  if
one requires de~Sitter invariance.  As a result, there is no de~Sitter-invariant Hadamard state for free minimally coupled massless scalar field
theory~\cite{Allen1985}.
Thus, there appears to be no de~Sitter-invariant perturbative vacuum state for the FP-ghost fields.  If this were indeed
the case, then
de~Sitter invariance would be broken via the FP-ghost sector in covariantly quantized Yang-Mills theory.
In Ref.~\cite{Faizal2008} it was pointed out that, if one regularizes the FP-ghost propagator by introducing a small mass, then the IR-divergent constant
term in the propagator does not contribute to the amplitude because the FP ghosts interact with the gauge field through a derivative coupling.  With this
observation, it was proposed
to use the effective IR-finite propagator defined by discarding the constant IR-divergent term and then taking the massless limit. This effective propagator
is de~Sitter-invariant.  However, the mass term for the FP ghosts breaks BRST invariance, and it is not entirely clear whether BRST
invariance is restored in the massless limit.

In this paper we present a new method to solve this IR problem in the FP-ghost sector in global de~Sitter space, which has compact spatial sections.
The resulting theory is compatible with BRST invariance of Yang-Mills theory.
The crucial observation in our method is that the spatially constant modes causing IR divergences,
which we call the zero modes, of the FP ghosts can be removed from the theory
by requiring the physical states to be
annihilated by certain conserved charges.  We explain our method in a wider class of spacetimes
with compact spatial sections to which global de~Sitter space
belongs. We also show that the theory we obtain is equivalent to that proposed in Ref.~\cite{Faizal2008}.

The rest of the paper is organized as follows.  In
Sec.~\ref{sec:IR-in-de-Sitter} we review the incompatibility between de~Sitter invariance and IR finiteness for the free minimally coupled massless scalar
field.  We then show that the zero modes can be removed from the theory by imposing the condition that
a certain conserved charge annihilate the physical states.
In Sec.~\ref{sec:conserved-charges} we discuss the conserved charges involving the FP ghosts
that will be used to remove the zero modes.  In particular, we discuss their relationship to the BRST and antiBRST charges.
In Sec.~\ref{sec:effective-Hamiltonian} we write down the Hamiltonian of the theory that governs the physical states that are annihilated by the conserved
charges discussed in Sec.~\ref{sec:conserved-charges} in a class of spacetimes that includes global de~Sitter space.
This effective Hamiltonian contains no zero modes but is coordinate dependent and has a nonlocal interaction term.
In Sec.~\ref{sec:comparison} we show that the theory described by the Hamiltonian found in Sec.~\ref{sec:effective-Hamiltonian}
is equivalent to that of Ref.~\cite{Faizal2008} obtained by discarding the
IR-divergent part of the FP-ghost propagator in the de~Sitter case.
Then we summarize and discuss our results in Sec.~\ref{sec:summary}.
In Appendix~\ref{appendix:deSitterzeromode} we provide some technical details for Sec.~\ref{sec:IR-in-de-Sitter}.
In Appendix~\ref{appendix:more-general-case} we present the effective Hamiltonian of Sec.~\ref{sec:effective-Hamiltonian}
for the general spacetime with
compact spatial sections.  In Appendix~\ref{appendix:dirac-bracket} we justify the identification of the
time component of the gauge field as (a multiple of) the conjugate momentum density for the Nakanishi-Lautrup auxiliary field by using the Dirac bracket.
In Appendix~\ref{appendix:group-averaging} we discuss the redefinition of the inner product necessary for defining the states annihilated by the conserved
charges of Sec.~\ref{sec:conserved-charges}.  In Appendix~\ref{appendix:flat-torus} we illustrate some of our results in the special case of
static torus space.  Finally, in Appendix~\ref{appendix:zero-mode-for-A} we show that the perturbative vacuum state for the de~Sitter case
is automatically annihilated by the bosonic Noether charges discussed in Sec.~\ref{sec:conserved-charges}.
We use natural units with $\hbar=c=1$ and the metric signature $+-\cdots-$ throughout this paper.

\section{Minimally coupled massless scalar field in de~Sitter space}\label{sec:IR-in-de-Sitter}

The problem we wish to discuss in this paper stems from the fact that the FP ghosts are minimally coupled massless scalar fields.  For this reason we first briefly
review some aspects of
the IR problem for the (free) minimally coupled scalar field theory and the nonexistence of de~Sitter-invariant Hadamard state for this theory~\cite{Allen1985}.   We then
show, as pointed out in Ref.~\cite{GarrigaKirsten},
how a de~Sitter-invariant state can be obtained in a new Hilbert space with a slightly modified
inner product.  (It is also known that there is a unitary
representation of the de~Sitter group corresponding to the minimally coupled massless scalar field~\cite{HiguchiJMP}.)
 This construction serves as a toy model
for what we shall do for the FP-ghost sector of Yang-Mills theory.

The line element of $n$-dimensional de~Sitter space in global coordinates is
\be
ds^2 = dt^2 -  H^{-2}\cosh^2 Ht\, d\Omega^2,
\ee
where $d\Omega^2$ is the line element of the $(n-1)$-dimensional unit sphere $S^{n-1}$ and where $H$ is a positive constant.  We assume $n\geq 3$ here.
Let $Y_{\ell\sigma}(\mbox{\boldmath$\theta$})$, $\ell=0,1,2,\ldots$, be the scalar spherical harmonics on $S^{n-1}$,
where $\mbox{\boldmath$\theta$}=(\theta_1,\theta_2,\ldots,\theta_{n-1})$
are the angular coordinates on $S^{n-1}$ and where $\sigma$ represents all labels other than $\ell$. They satisfy
\be
\eta^{ij}\tilde{\nabla}_i \tilde{\nabla}_j Y_{\ell\sigma}(\mbox{\boldmath$\theta$})
= - \ell(\ell+n-2)Y_{\ell\sigma}(\mbox{\boldmath$\theta$}),
\ee
where $\tilde{\nabla}_i$ is the covariant derivative on $S^{n-1}$ with metric $\eta_{ij}$ and inverse metric $\eta^{ij}$.  The spherical harmonics are
normalized as
\be
\int_{S^{n-1}} d\Omega\, Y_{\ell\sigma}^*(\mbox{\boldmath$\theta$}) Y_{\ell'\sigma'}(\mbox{\boldmath$\theta$}) = \delta_{\ell\ell'}\delta_{\sigma\sigma'}.
\ee
Then, the minimally coupled  scalar field $\phi(t,\mbox{\boldmath$\theta$})$
of mass $M$  can be expanded as
\be
\phi(t,\mbox{\boldmath$\theta$})
= \sum_{\ell=0}^\infty \sum_\sigma
\left[ a_{\ell\sigma}f_\ell(t)Y_{\ell\sigma}(\mbox{\boldmath$\theta$})
+ a_{\ell\sigma}^\dagger f^*_\ell(t) Y_{\ell\sigma}^*(\mbox{\boldmath$\theta$})\right].
\ee
 The functions $f_\ell(t)$ are given by
\be
f_\ell(t) =H^{\frac{n-2}{2}} N_{\ell} (\cosh Ht)^{-\frac{n-2}{2}}\mathrm{P}_{-\frac{1}{2} + \lambda}^{-\ell - \frac{n-2}{2}}(i \sinh Ht),
\label{original}
\ee
where
\bea
\lambda & = & \sqrt{\left(\frac{n-1}{2}\right)^2 - \left(\frac{M}{H}\right)^2},\\
|N_{\ell}|^2 & = & \frac{\Gamma(\ell+\frac{n-1}{2} + \lambda)\Gamma(\ell + \frac{n-1}{2} - \lambda)}{2}. \label{Nell}
\eea
(See, e.g., Ref.~\cite{HiguchiJMP}.)  The associated Legendre function is given by
\bea
\mathrm{P}_\nu^{-\mu}(x) & = & \frac{1}{\Gamma(1+\mu)}\left(\frac{1-x}{1+x}\right)^{\frac{\mu}{2}} \nonumber \\
&& \times F\left(-\nu, \nu+1;1+\mu; \frac{1-x}{2}\right),
\eea
where $F(\alpha,\beta;\gamma;z)$ is Gauss's hypergeometric function.
The normalization constant $N_\ell$ in Eq.~\eqref{Nell} has been determined by the requirement
 \bea
\left[a_{\ell\sigma},a^\dagger_{\ell'\sigma'}\right] & = & \delta_{\ell\ell'}\delta_{\sigma\sigma'},\\
\left[ a_{\ell\sigma},a_{\ell'\sigma'}\right] & = & 0,
\eea
and the equal-time commutation relations of
the field operator $\phi(t,\mbox{\boldmath$\theta$})$ and its time derivative.  In particular,
\be
\left[\phi(t,\mbox{\boldmath$\theta$}),\dot{\phi}(t,\mbox{\boldmath$\theta$}')\right] =
\frac{i H^{n-1}}{\cosh^{n-1} Ht}\delta(\mbox{\boldmath$\theta$},\mbox{\boldmath$\theta$}'). \label{equal-time-com}
\ee
 Equation~\eqref{original} can be rewritten by using
\be
F(\alpha,\beta;\gamma;z) = (1-z)^{\gamma-\alpha-\beta}F(\gamma-\alpha,\gamma-\beta;\gamma;z) \label{elementary}
\ee
as
\bea
f_\ell(t) & = & \frac{H^{\frac{n-2}{2}}N_\ell(\cosh Ht)^{\ell}}{2^{\ell+\frac{n-2}{2}} \Gamma(\ell+\frac{n}{2})} \nonumber \\
&& \times F\left(b_{\ell+},b_{\ell-};\ell+\frac{n}{2};\frac{1-i\sinh Ht}{2}\right), \label{secondary}
\eea
where
\be
b_{\ell\pm} = \ell+\frac{n-1}{2}\pm \lambda.  \label{b-ell-pm}
\ee

The de~Sitter-invariant Bunch-Davies (or Euclidean)
vacuum state $|0\rangle$~\cite{ChernikovTagirov,BunchDavies,GibbonsHawking} is defined by requiring
$a_{\ell\sigma}|0\rangle = 0$ for all $\ell$ and $\sigma$.  The two-point Wightman function in this state is
\be
\langle 0|\phi(t,\mbox{\boldmath$\theta$})\phi(t',\mbox{\boldmath$\theta$}')|0\rangle
= \sum_{\ell=0}^\infty
f_\ell(t)f_\ell^*(t') \sum_{\sigma} Y_{\ell\sigma}(\mbox{\boldmath$\theta$})Y_{\ell\sigma}^*(\mbox{\boldmath$\theta$}'). \label{phi-expand}
\ee
This two-point function is divergent as $M\to 0^+$ because the space-independent mode function,
\be
F_0(t) := f_0(t) Y_{00}(\mbox{\boldmath$\theta$}),
\ee
where $Y_{00}(\mbox{\boldmath$\theta$})$ is the space-independent spherical harmonic with $\ell=0$,
is divergent in this limit.  Hence, the two-point function given by Eq.~\eqref{phi-expand} is divergent and
the Bunch-Davies vacuum state will not be well defined in the massless limit.

In Appendix~\ref{appendix:deSitterzeromode} it is shown that
\be
F_0(t) =\frac{1}{\sqrt{2c_0}} \left\{ \frac{1}{M}-  M\left[ g(t) + c_1 + i c_0 f(t)\right]\right\} + o(M), \label{answerF}
\ee
where
\bea
c_0 & = & \frac{\pi^{(n+1)/2}}{\Gamma(\frac{n+1}{2})H^{n}},  \label{c0def}\\
f(t) & = & \int_0^t \frac{dt'}{V(t')}, \label{ft}\\
g(t) & = & \int_0^t dt'\left[\frac{1}{V(t')}\int_0^{t^{\prime}} dt^{\prime\prime}V(t^{\prime\prime})\right]. \label{gt}
\eea
Here,
\be
V(t) : = \frac{2\pi^{n/2}}{\Gamma(\frac{n}{2})H^{n-1}}\cosh^{n-1} Ht \label{Vt-formula}
\ee
is the volume of the spatial section at time $t$, which is an $(n-1)$-dimensional sphere of radius $H^{-1}\cosh Ht$.
Note that
\be
V_{S^{n-1}} = \frac{2\pi^{n/2}}{\Gamma(\frac{n}{2})} \label{V-unit-sphere}
\ee
is the volume of the unit $S^{n-1}$.
We do not need the value of the constant $c_1$, which is given in Eq.~\eqref{c-1-value} for completeness.

Although there is no de~Sitter-invariant Bunch-Davies vacuum for the minimally coupled massless scalar field, there are $O(n)$-invariant
states~\cite{Allen1985,AllenFolacci}.  To see this, we note first that the $\ell=0$ sector of the field $\phi(t,\mbox{\boldmath$\theta$})$, i.e.\ the
space-independent sector, which we denote by $\phi_{(0)}(t)$ and call the zero mode, satisfies
\be
\frac{d\ }{dt}\left[ V(t)\frac{d\phi_{(0)}(t)}{dt}\right] = 0,
\ee
with the general solution
\be
\phi_{(0)}(t) = \hat{q} + \hat{p}f(t),
\ee
where $f(t)$ is defined by Eq.~\eqref{ft}.  Thus, the massless field $\phi(t,\mbox{\boldmath$\theta$})$ can be written, with the zero mode
separated out, as
\bea
\phi(t,\mbox{\boldmath$\theta$})
& = & \hat{q} + \hat{p}f(t) \nonumber \\
&& +  \sum_{\ell=1}^\infty \sum_\sigma
\left[ a_{\ell\sigma}f_\ell(t)Y_{\ell\sigma}(\mbox{\boldmath$\theta$})
+ a_{\ell\sigma}^\dagger f^*_\ell(t) Y_{\ell\sigma}^*(\mbox{\boldmath$\theta$})\right]. \nonumber \\
\label{field-expansion}
\eea
By integrating the equal-time commutator~\eqref{equal-time-com} over the space we find
\be
[\hat{q},\hat{p}] = i,
\ee
which is identical with the commutator between the position and momentum operators in one-dimensional quantum mechanics.  Thus, one can represent the
zero-mode sector by a normalized
wave function $\psi(q)$ and let $\hat{p} = -i d/dq$.  If we define $|0_{(+)}\rangle$ to be the vacuum state for the $\ell\neq 0$ sector satisfying
the requirement $a_{\ell\sigma}|0_{(+)}\rangle = 0$ for all $\ell> 0$ and $\sigma$, then the state
$|\psi\rangle = \psi(q)\otimes |0_{(+)}\rangle$ is $O(n)$ invariant but not de~Sitter-invariant.

To show that the state $|\psi\rangle$ is not de~Sitter-invariant, we examine the action of a de~Sitter boost  on the
operators $\hat{q}$, $\hat{p}$ and $a_{\ell\sigma}$.
Let us parametrize the unit $S^{n-1}$ using the natural embedding space $\mathbb{R}^{n}$ as
$x_1 = \cos\chi$, $x_2 =  \tilde{x}_2\sin\chi$, ..., $x_n =  \tilde{x}_n\sin \chi$,
where $\chi \in [0,\pi]$ and
$\tilde{x}_2^2 +\cdots + \tilde{x}_n^2 = 1$.
Let $\tilde{Y}_{m\tilde{\sigma}}$ be the spherical harmonics on the unit $S^{n-2}$ with eigenvalues $-m(m+n-3)$, $m=0,1,2,\ldots$, of the Laplacian there.
Then the spherical harmonics $Y_{\ell m\tilde{\sigma}}$ on $S^{n-1}$ proportional to $\tilde{Y}_{m\tilde{\sigma}}$
are given by~\cite{HiguchiJMP}
\be
Y_{\ell m\tilde{\sigma}}(\mbox{\boldmath$\theta$}) = c_{\ell m}(\sin\chi)^{-(n-3)/2}\mathrm{P}_{\ell+(n-3)/2}^{-(m+(n-3)/2)}(\cos\chi)
\tilde{Y}_{m\tilde{\sigma}},
\ee
where
\be
c_{\ell m} = \left[ \frac{2\ell+n-2}{2}\cdot \frac{(\ell+m+n-2)!}{(\ell-m)!}\right]^{1/2}.
\ee
It is clear that the mode functions given by
\be
F_{\ell m \tilde{\sigma}}(t,\mbox{\boldmath$\theta$}) = f_\ell(t) Y_{\ell m\tilde{\sigma}}(\mbox{\boldmath$\theta$})
\ee
transform
to one another, i.e. they do not transform into $F^*_{\ell m \tilde{\sigma}}(t,\mbox{\boldmath$\theta$})$,
under the $O(n)$ transformations.  If $M>0$, it can
be shown that the functions $F_{\ell m \tilde{\sigma}}(t,\mbox{\boldmath$\theta$})$ transform among themselves under de~Sitter boosts
as well~\cite{HiguchiJMP}.
To show this, it is sufficient to examine the action of the boost Killing vector in the $x_1$ direction given by
\be
L_X : = \frac{\cos\chi}{H}\frac{\partial\ }{\partial t} - \tanh Ht \sin \chi\frac{\partial\ }{\partial\chi}, \label{LX-def}
\ee
because the component connected to the identity of the de~Sitter group is generated by $L_X$ and the $SO(n)$ rotations.
Indeed we show in Appendix~\ref{appendix:deSitterzeromode} that
\bea
L_X F_{\ell m \tilde{\sigma}}(t,\mbox{\boldmath$\theta$})
& = &  -i k_{\ell m} F_{(\ell-1)m\tilde{\sigma}}(t,\mbox{\boldmath$\theta$})\nonumber \\
&& -i k_{(\ell+1) m} F_{(\ell+1)m\tilde{\sigma}}(t,\mbox{\boldmath$\theta$}), \label{LX-eq}
\eea
where
\bea
k_{\ell m} & = & \left[\frac{(\ell - m)(\ell+m+n-3)}{(2\ell+n-2)(2\ell+n-4)}\right]^{1/2} \nonumber \\
&& \times \left[ (\ell-1)(\ell+n-2) + \frac{M^2}{H^2}\right]^{1/2}.
\eea

The transformation~\eqref{LX-eq} is unaltered for the massless case if $\ell\geq 2$ or $(\ell,m)= (1,1)$, but for $(\ell,m)=(0,0)$ and $(\ell,m)=(1,0)$
it needs to be reexamined because
the $\ell=0$ mode function $F_0(t)$ is divergent in the massless limit.
Using Eq.~\eqref{answerF} in Eq.~\eqref{LX-eq} with $\ell=1$ and $m=0$, we obtain
\be
L_X F_{10}(t,\mbox{\boldmath$\theta$})
 =  - \frac{i}{\sqrt{2nc_0 H^2}} - i k_{20}  F_{20}(t,\mbox{\boldmath$\theta$}).
\ee
We have omitted the label $\tilde{\sigma}$ for $m=0$ because there is only one independent spherical harmonic for each $\ell$ with $m=0$.
 Next we observe that
\bea
L_X f(t) & =  & - \sqrt{\frac{2}{c_0}}\lim_{M\to 0^+}M^{-1}\textrm{Im}[L_X F_0(t)] \nonumber \\
& = & \frac{1}{\sqrt{2nc_0 H^2}}\left[ F_{10}(t,\mbox{\boldmath$\theta$}) + F^*_{10}(t,\mbox{\boldmath$\theta$})\right].
\eea

The boost $L_X$ acts on the field $\phi(t,\mbox{\boldmath$\theta$})$ given by Eq.~(\ref{field-expansion})  as
\bea
&& L_X \phi(t,\mbox{\boldmath$\theta$})\nonumber \\
&&  =   \hat{p}L_X f(t)\nonumber \\
&&  \,\,\, +  \sum_{\ell=1}^\infty\sum_{m=0}^\ell \sum_{\tilde{\sigma}}
\left[ a_{\ell m\tilde{\sigma}}L_X F_{\ell m\tilde{\sigma}}(t,\mbox{\boldmath$\theta$})
+ a_{\ell m \tilde{\sigma}}^\dagger L_X  F^*_{\ell m\tilde{\sigma}}(t,\mbox{\boldmath$\theta$})\right]. \nonumber \\
\eea
We rewrite this boost transformation of $\phi(t,\mbox{\boldmath$\theta$})$ in such a way that the transformation is attributed to the operators:
\bea
&& L_X \phi(t,\mbox{\boldmath$\theta$})\nonumber \\
&&  =   \delta_X \hat{q} +  f(t)\delta_X \hat{p}\nonumber \\
&&  \,\,\, +  \sum_{\ell=1}^\infty\sum_{m=0}^\ell \sum_{\tilde{\sigma}}
\left[  F_{\ell m\tilde{\sigma}}(t,\mbox{\boldmath$\theta$})\delta_X a_{\ell m\tilde{\sigma}}
+ F^*_{\ell m\tilde{\sigma}}(t,\mbox{\boldmath$\theta$})\delta_X a_{\ell m \tilde{\sigma}}^\dagger  \right]. \nonumber \\
\eea
Thus, for $\ell\geq 2$ and $(\ell,m) = (1,1)$ we find
\be
\delta_X a_{\ell m\tilde{\sigma}}  = - i k_{\ell m} a_{(\ell - 1)m\tilde{\sigma}} - ik_{(\ell+1)m}a_{(\ell+1)m\tilde{\sigma}}.
\ee
For the other operators we have
\bea
\delta_X \hat{q} & = & \frac{i}{\sqrt{2nc_0 H^2}}(a_{10}^\dagger - a_{10}),\\
\delta_X \hat{p} & = & 0, \label{pX}\\
\delta_X a_{10} & = & \frac{1}{\sqrt{2nc_0 H^2}}\hat{p} - ik_{20} a_{20}. \label{a100}
\eea
(The invariance of $\hat{p}$ under de~Sitter transformations
can also be inferred by noting that $\hat{p}$ is the Noether charge corresponding to the conserved current $\nabla^\mu \phi$
of the theory.)

Equation~\eqref{a100} implies that the conditions $a_{\ell m \tilde{\sigma}}|\psi\rangle = 0$ are not de~Sitter-invariant unless the condition
$\hat{p}|\psi\rangle= 0$ is also imposed.  Conversely, Eq.~\eqref{pX} implies that these conditions taken together are de~Sitter-invariant.
Thus, the vacuum state $|\psi\rangle$ defined by requiring $a_{\ell\sigma}|\psi\rangle = 0$ and $\hat{p}|\psi\rangle =0$ is de~Sitter-invariant.  However,
the condition $\hat{p}|\psi\rangle = - i\psi'(q)\otimes |0_{(+)}\rangle = 0$  implies that $\psi(q)$ is a constant function.  Then $\int dq|\psi(q)|^2 = \infty$.
Therefore, there is no normalizable de~Sitter-invariant state of the form $\psi(q)\otimes |0_{(+)}\rangle$.

However, it is possible to redefine the inner product of states $|\Psi_1\rangle = \psi(q)\otimes |\alpha_1\rangle$ and
$|\Psi_2\rangle = \psi(q)\otimes |\alpha_2\rangle$, where $|\alpha_1\rangle$ and $|\alpha_2\rangle$ are states in the Fock space built
by applying creation operators $a_{\ell \sigma}$ on $|0_{(+)}\rangle$ and where $\psi(q)$ is constant, simply as
$\langle \Psi_1|\Psi_2\rangle = \langle\alpha_1|\alpha_2\rangle$.  The state $\psi(q)\otimes |0_{(+)}\rangle$ with $\psi(q)$ = const.\ is a well-defined
de~Sitter-invariant state with this inner product.
A similar redefinition of the inner product was used in quantum cosmology~\cite{Marolf1995}.  This redefinition is closely related
to the method of ``group averaging'' (see, e.g.\ Ref.~\cite{Higuchi1991a,Higuchi1991b,LandsmanGA}), which has been incorporated into the refined
algebraic quantization~\cite{Ashtekar1995}.

If the operators $\hat{q}$ and $\hat{p}$ are physical observables,
this redefinition of inner product described above will not be physical because, e.g.\ the expectation
value of $\hat{q}$ will be undefined.  However, since the
FP ghosts are not physical particles, a similar redefinition of inner product for these fields
will not affect true physical quantities.  In the next section we identify conserved
charges analogous to the operator $\hat{p}$ that can be used to `banish' the zero modes of the FP ghosts.

\section{Some conserved charges in covariantly quantized Yang-Mills theory}\label{sec:conserved-charges}

In this section we examine covariantly quantized Yang-Mills theory in the Landau gauge in (globally hyperbolic)
spacetime with compact spatial sections.  Let $f^{abc}$ be
the totally antisymmetric structure constant for a compact semisimple Lie group. Thus, if $T^a$ are the generators of this group, then
\be
[T^a,T^b] = if^{abc}T^c,
\ee
where the repeated Lie-algebra indices are summed over.
The fields in Yang-Mills theory are the gauge field $A^a_\mu$ and the FP-ghost and antighost fields, $c^a$ and $\overline{c}^a$.  It is possible to introduce
other gauge multiplets, but we do not do so in this paper.  One defines the field strength as
$F^a_{\mu\nu} = \nabla_\mu A_\nu^a - \nabla_\nu A_\mu^a + q f^{abc}A^b_{\mu}A^c_{\nu}$, where $q$ is the coupling constant.
We introduce the following notation~\cite{KugoOjima1979}:
\bea
X\cdot Y & :  = & X^a Y^a,\\
(X\times Y)^a &  : = & f^{abc}X^b X^c.
\eea
In this simplified notation the field strength is given as
\be
F_{\mu\nu} = \nabla_\mu A_\nu - \nabla_\nu A_\mu + q A_\mu \times A_\nu.
\ee

The Lagrangian density is given by
\bea
\mathcal{L} & = & \sqrt{|g|}
\left\{ - \frac{1}{4}F_{\mu\nu}\cdot F^{\mu\nu}  - i \nabla^\mu \overline{c}\cdot D_\mu c\right. \nonumber \\
&& \,\,\,\,\,\,\,\, \left. - \frac{1}{2\xi}\nabla_\mu A^\mu\cdot \nabla_\nu A^\nu\right\},  \label{lagrangian1}
\eea
where
\be
D_\mu c = \nabla_\mu c + q A_\mu \times c.
\ee
Here, $g$ is the determinant of the background spacetime metric $g_{\mu\nu}$, and the FP-ghost field $c^a$ and
antighost field $\overline{c}^a$ are fermionic Hermitian
fields~\cite{KugoOjima1978,KugoOjima1979}.  This Lagrangian density can be rewritten by introducing the Nakanishi-Lautrup auxiliary field
$B^a$~\cite{Nakanishi,Lautrup} as
\bea
\mathcal{L} & = & \sqrt{|g|}
\left\{ - \frac{1}{4}F_{\mu\nu}\cdot F^{\mu\nu}  - i\nabla^\mu\overline{c}\cdot D_\mu c \right. \nonumber \\
&&  \left.  - \nabla_\mu B \cdot A^\mu  + \frac{\xi}{2}B\cdot B\right\}.  \label{lagrangian2}
\eea
One can eliminate the field $B^a$ using its field equation from this Lagrangian density
and show that it is equivalent to the Lagrangian density in Eq.~\eqref{lagrangian1} up to
a total derivative.

The Lagrangian density~\eqref{lagrangian2} is invariant (up to a total derivative) under the BRST transformation~\cite{BRS,Tuytin}:
\bea
\delta_B A_\mu & = & \epsilon D_\mu c,\\
\delta_B c & = & - \tfrac{1}{2}\epsilon q  c\times c,\\
\delta_B \overline{c} & = & i\epsilon B,\\
\delta_B B & = & 0,
\eea
where $\epsilon$ is a Grassmann number anticommuting with $c$ and $\overline{c}$.  The conserved current corresponding to this invariance can be given
as~\cite{KugoOjima1979}
\be
J_B^\mu = B\cdot D^\mu c - \nabla^\mu B\cdot c + \tfrac{i}{2}q \nabla^\mu\overline{c}\cdot (c\times c) - \nabla_\nu (F^{\mu\nu}\cdot c).
\ee
The conserved BRST charge is
\be
Q_B = \int_{\Sigma}d\Sigma_\mu J_B^\mu,
\ee
where $\Sigma$ is a Cauchy surface.   In this paper we specialize to the Landau gauge $\xi = 0$.  Then the Lagrangian density~\eqref{lagrangian2} is
\be
\mathcal{L}_{\textrm{Lan}} = \sqrt{|g|}
\left\{ - \frac{1}{4}F_{\mu\nu}\cdot F^{\mu\nu}  - i\nabla^\mu\overline{c}\cdot D_\mu c - \nabla_\mu B \cdot A^\mu \right\}.
\label{lagrangian3}
\ee

The FP and gauge-fixing terms in the action can be rewritten as follows:
\bea
&& \int dx \sqrt{|g|} \left\{ -i\nabla^\mu \overline{c}\cdot D_\mu c - \nabla_\mu B\cdot A^\mu\right\} \nonumber \\
&& =\int dx \sqrt{|g|} \left\{ i\nabla^\mu c \cdot D_\mu \overline{c} - \nabla_\mu (B - iq \overline{c}\times c)\cdot A^\mu\right\}, \nonumber \\
\eea
where the integral is over the spacetime and where the surface terms have been dropped. The second form of this part of the action can be obtained
from the first by interchanging $c$ and $\overline{c}$ and changing $B$ to $B-iq\overline{c}\times c$.  This
makes it clear that the Lagrangian density $\mathcal{L}_{\textrm{Lan}}$ is also invariant (up to a total derivative)
under the following transformation called the antiBRST transformation~\cite{CurciFerrari,Ojima1980,Baulieu1982}:
\bea
\delta_{\overline{B}} A_\mu & = & \epsilon D_\mu \overline{c},\\
\delta_{\overline{B}} \overline{c} & = & - \tfrac{1}{2}\epsilon q  \overline{c}\times \overline{c},\\
\delta_{\overline{B}} c & = & - i\epsilon (B- iq \overline{c}\times c),\\
\delta_{\overline{B}} B & = & - q\epsilon\overline{c}\times B.
\eea
(In fact the antiBRST transformation can be defined for all values of $\xi$~\cite{Ojima1980}.)

Now, the Euler-Lagrange equations of motion arising from the variations of $B$, $c$ and $\overline{c}$ are
\bea
\nabla_\mu A^\mu & = & 0, \label{divergenceA}\\
\nabla_\mu D^\mu c & = & 0, \label{c-eq}\\
\nabla_\mu D^\mu\overline{c} & = & 0. \label{c-bar-eq}
\eea
One needs to use Eq.~\eqref{divergenceA} to derive Eq.~\eqref{c-bar-eq} from the original Euler-Lagrange equation.
If we let $q=0$ in the last two equations, we find $\nabla_\mu \nabla^\mu c = \nabla_\mu\nabla^\mu\overline{c} = 0$.  That is, the FP ghost and antighost
are minimally coupled massless scalar fields.

Eqs.~\eqref{divergenceA}-\eqref{c-bar-eq} imply that the following charges are conserved:
\bea
Q_A & = & \int_\Sigma d\Sigma_\mu A^\mu, \label{Q_A-def}\\
Q_{Dc} & = & \int_\Sigma d\Sigma_\mu D^\mu c,\\
Q_{D\bar{c}} & = & \int_\Sigma d\Sigma_\mu D^\mu \overline{c},
\eea
where $\Sigma$ is a Cauchy surface. (Note that these charges carry a Lie-algebra index.)
These are the Noether charges corresponding to symmetries of the Lagrangian, as can readily be verified.
Our proposal is to require that the physical states be annihilated by these charges.
Thus, we impose the following conditions on the physical states $|\textrm{phys}\rangle$ as well as the usual condition $Q_B|\textrm{phys}\rangle=0$:
\be
Q_A|\textrm{phys}\rangle = Q_{Dc}|\textrm{phys}\rangle = Q_{D\bar{c}}|\textrm{phys}\rangle = 0.  \label{the-conditions}
\ee
The symmetries these charges generate are spacetime scalars.  Therefore, they commute with spacetime symmetry generators.  As a result,
the conditions~\eqref{the-conditions} are invariant under any continuous spacetime symmetries.
As we shall see in the next section, these conditions eliminate the spatially constant modes from FP ghosts.  For the de~Sitter case this will lead to
de~Sitter-invariant perturbation theory.

In fact, the condition $Q_{Dc}|\textrm{phys}\rangle = 0$ is a consequence of the condition $Q_A|\textrm{phys}\rangle = 0$ and the BRST invariance of
the physical states because $[Q_B,Q_A] = -i Q_{Dc}$~\cite{KugoOjima1979}.  If we require antiBRST invariance, with the corresponding conserved charge
$Q_{\overline{B}}$, of the physical states as well, then the
condition
$Q_{D\bar{c}}|\textrm{phys}\rangle = 0$ will also be a consequence of the condition $Q_A|\textrm{phys}\rangle = 0$ because
$[ Q_{\overline{B}}, Q_A] = -i Q_{D\bar{c}}$.
These observations naturally lead to the observation that one also needs to impose the condition
$\{Q_B,Q_{D\overline{c}}\}|\textrm{phys}\rangle =0$. (Note that $\{Q_B,Q_{Dc}\} =  0$ because of the nilpotency of $Q_B$, i.e.\ $Q_B^2 = 0$.)
This condition turns out to have a natural interpretation.  The field equation for $A_\mu$ can be written as~\cite{OjimaNP,Ojima1980}
\be
\nabla^\nu F_{\nu \mu} + q J_\mu = \{Q_B,D_\mu\overline{c}\}  = - \{Q_{\overline{B}},D_\mu c\},  \label{global-gauge}
\ee
where $J^\mu$ is the Noether current for the global gauge transformation~\cite{KugoOjima1979}:
\be
J^\mu  : =  A_\nu \times F^{\nu\mu} + A^\mu \times B - i\,\overline{c}\times D^\mu c + i\, \nabla^\mu \overline{c} \times c.
\ee
The corresponding Noether charge is
\be
Q_{\textrm{gg}} = \int_\Sigma d\Sigma_\mu J^\mu.
\ee
By integrating Eq.~\eqref{global-gauge} over a (compact) Cauchy surface, we obtain
\be
qQ_{\textrm{gg}}  = \{ Q_B, Q_{D\bar{c}}\} = - \{Q_{\overline{B}},Q_{Dc}\}.
\ee
Thus, the condition  $Q_{\textrm{gg}}|\textrm{phys}\rangle = 0$
on the physical states results from the condition $Q_{D\bar{c}}|\textrm{phys}\rangle = 0$ and
the BRST invariance of the physical states.  It will also result from the condition $Q_{Dc}|\textrm{phys}\rangle = 0$ if we require the antiBRST invariance of the
physical states.

The nilpotency of the BRST and antiBRST charges imply that $\{Q_B,Q_{\textrm{gg}}\} = \{Q_{\overline{B}},Q_{\textrm{gg}}\} = 0$.  Thus,
the conditions given by Eq.~\eqref{the-conditions} are compatible with the BRST and antiBRST invariance if we also impose the condition
$Q_{\textrm{gg}}|\textrm{phys}\rangle = 0$, i.e.\ the requirement that the physical state be invariant under the global gauge transformations.
No more conditions are required for consistency of our conditions~\eqref{the-conditions} with BRST or antiBRST invariance.
(One can show that $\{Q_B,Q_{\overline{B}}\} = 0$ as is well known and that $\{ Q_{Dc}^a,Q_{D\bar{c}}^b\} = qf^{abc}Q_A^c$.)
In the next section we show that the conditions~\eqref{the-conditions} lead to an effective Hamiltonian that does not depend on the zero modes of the FP ghosts.

\section{The Hamiltonian without zero modes}\label{sec:effective-Hamiltonian}

In this section we analyze the FP ghosts of Yang-Mills theory in spacetime with compact spatial sections with the following line element:
\be
ds^2 = [N(\mathbf{x})]^2 dt^2 - \gamma_{ij}(t,\mathbf{x}) dx^i dx^j, \label{our-metric}
\ee
where $\mathbf{x} = (x_1,x_2,\ldots, x_{n-1})$.  We also assume that the determinant of the spatial metric factorizes as
\be
\gamma(t,\mathbf{x}) = h_1(t)h_2(\mathbf{x}).  \label{gamma-factorization}
\ee
The standard metric of global de~Sitter space satisfies this property.
 We shall show that the conditions~\eqref{the-conditions} lead to a Hamiltonian without zero modes.
We discuss the Hamiltonian with a line element of a more general form
in Appendix~\ref{appendix:more-general-case}.

The Lagrangian is obtained by integrating the Lagrangian density~\eqref{lagrangian3} over a Cauchy surface of constant time as
\be
L = \int d\mathbf{x}\,\mathcal{L}_{\textrm{Lan}}.
\ee
The conjugate momentum densities for the fields $A_i$, $B$, $c$ and $\overline{c}$ can be found as follows:
\bea
\pi_{A}^i(t,\mathbf{x})  & = & \frac{\delta L}{\delta \dot{A}_i (t,\mathbf{x})} = - \sqrt{|g|} F^{0i}(t,\mathbf{x}),\\
\pi_B (t,\mathbf{x}) & = & \frac{\delta L}{\delta \dot{B}(t,\mathbf{x})} = - \sqrt{|g|} A^0(t,\mathbf{x}), \label{pi-B} \\
\pi_c(t,\mathbf{x}) & = & \frac{\delta L}{\delta \dot{c}(t,\mathbf{x})} = i\sqrt{|g|}  \nabla^0  \overline{c}(t,\mathbf{x}), \label{pi-c}\\
\pi_{\bar{c}}(t,\mathbf{x}) & = & \frac{\delta L}{\delta \dot{\overline{c}}(t,\mathbf{x})}  = -  i\sqrt{|g|} D^0 c(t,\mathbf{x}), \label{pi-c-bar}
\eea
where the functional derivative with respect to a fermionic variable, e.g.\ $\delta L/\delta\dot{c}(t,\mathbf{x})$, is taken from the left, and where the
indices are raised by the full inverse metric $g^{\mu\nu}$.
 Equation~\eqref{pi-B} is a second-class constraint.  We show in Appendix~\ref{appendix:dirac-bracket} that
the quantization using the Dirac bracket to deal with this constraint equation is equivalent to regarding
it as defining $-\sqrt{|g|}A^0:=\pi_B$ without treating $A_0^a$ as independent dynamical variables.

The Hamiltonian density can readily be found as follows:
\bea
\mathcal{H} & = & \dot{A}_i \cdot \pi^i_A + \dot{B}\cdot \pi_B + \dot{c}\cdot \pi_c + \dot{\overline{c}}\cdot \pi_{\bar{c}} - \mathcal{L}_{\textrm{Lan}}
\nonumber \\
& = & \mathcal{H}_{\textrm{class}} + \mathcal{H}_{\textrm{GF}+\textrm{FP}},
\eeq
where the classical contribution to the Hamiltonian density is
\bea
\mathcal{H}_{\textrm{class}} & = & \frac{N}{2\sqrt{\gamma}}\gamma_{ij}\pi^i\cdot \pi^j
+ \frac{\sqrt{\gamma}\, N}{4}\gamma^{ij}\gamma^{mn}F_{im}\cdot F_{jn} \nonumber \\
&& +  D_i A_0\cdot \pi^i,
\eea
with $A_0 = - N\pi_{B}/\sqrt{\gamma}$,
and where the contribution from the gauge-fixing and FP terms is
\bea
\mathcal{H}_{\textrm{GF}+\textrm{FP}}
& = & -\frac{N}{\sqrt{\gamma}}\pi_c \cdot \left( i \pi_{\bar{c}} + q \pi_{B} \times c\right)  \nonumber \\
&& - N\sqrt{\gamma}\,\gamma^{ij}\left( i \nabla_i \overline{c}\cdot D_j c +  \nabla_i B \cdot A_j\right)  \label{H-GF+FP-in-momentum}\\
& =  &  - i \frac{\sqrt{\gamma}}{N}\nabla_0\overline{c}\cdot \nabla_0 c \nonumber \\
&& - N\sqrt{\gamma}\,\gamma^{ij}\left(  i \nabla_i \overline{c}\cdot D_j c +  \nabla_i B \cdot A_j\right).  \label{H-GF+FP-before}
\eea
Here, the tensor $\gamma^{ij}$ is the inverse of $\gamma_{ij}$ as a matrix.

Now, let us define the zero mode for the field $c$ by
\be
c_{(0)}(t)  : = \frac{1}{V(t)}\int d\mathbf{x}\,\frac{\sqrt{\gamma}}{N} c(t,\mathbf{x}), \label{define-c0}
\ee
where
\be
V (t) : = \int d\mathbf{x} \frac{\sqrt{\gamma}}{N}. \label{vol-original}
\ee
Let us also define $c_{(+)}: = c - c_{(0)}$.  It is clear that
\be
\int d\mathbf{x}\,\frac{\sqrt{\gamma}}{N}\,c_{(+)}(t,\mathbf{x}) = 0.
\ee
We define $\overline{c}_{(0)}$ and $\overline{c}_{(+)}$ from the field $\overline{c}$  in the same way. Our aim is to construct a Hamiltonian that
does not depend on the
zero modes $c_{(0)}$ and $\overline{c}_{(0)}$ after imposing the conditions~\eqref{the-conditions}.

The Hamiltonian density
$\mathcal{H}_{\textrm{GF}+\textrm{FP}}$ given by Eq.~\eqref{H-GF+FP-in-momentum} depends on $c_{(0)}$ as it stands.  However,
this dependence can be eliminated by the following redefinition of the Nakanishi-Lautrup auxiliary field:
\be
\tilde{B} : = B - iq \overline{c}\times c_{(0)}. \label{field-redefinition}
\ee
We define the fields $\tilde{B}_{(0)}$ and $\tilde{B}_{(+)}$ for $\tilde{B}$ in the same way as those for $c$ and $\overline{c}$.
By substituting Eq.~\eqref{field-redefinition} into the Lagrangian density~\eqref{lagrangian3} we find that the new canonical conjugate
momentum densities of $\tilde{B}$, $c$ and $\overline{c}$ are
\bea
\Pi_{\tilde{B}} & = & \pi_{\tilde{B}} = - \sqrt{|g|} A^0 = \frac{\sqrt{\gamma}}{N}\varpi_{\tilde{B}}, \\
\Pi_{c}  & = &  \frac{\sqrt{\gamma}}{N}\varpi_c,\\
\Pi_{\bar{c}} &   = &   \frac{\sqrt{\gamma}}{N}\varpi_{\bar{c}},
\eea
where
\bea
\varpi_{\tilde{B}} & := & - A_0,\\
\varpi_{c} & : = & i\left[\nabla_0 \overline{c} - \frac{q}{V}\int d\mathbf{x} \frac{\sqrt{\gamma}}{N} \varpi_{\tilde{B}}
\times \overline{c}\right], \label{varpi-c}\\
 \varpi_{\bar{c}} & : = & -i\left[ \nabla_0 c - q \varpi_{\tilde{B}} \times c_{(+)}\right]. \label{varpi-bar-c}
\eea

We define the zero modes $\varpi_{\tilde{B}(0)}$ of $\varpi_{\tilde{B}}$ as
\be
\varpi_{\tilde{B}(0)}(t) : = \frac{1}{V}\int d\mathbf{x}\frac{\sqrt{\gamma}}{N} \varpi_{\tilde{B}}(t,\mathbf{x}).
\ee
The zero modes $\varpi_{c(0)}$ and $\varpi_{\bar{c}(0)}$ are defined in exactly the same way from $\varpi_{c}$ and $\varpi_{\bar{c}}$.
We also define
\be
\varpi_{X(+)} : = \varpi_{X} - \varpi_{X(0)},\,\,\, X = \tilde{B}, c, \bar{c}.
\ee
Then the equal-time canonical (anti)commutation relations, $[ \Pi_{X}(t,\mathbf{x}'), X(t,\mathbf{x})]_{\pm} =- i\delta(\mathbf{x},\mathbf{x}')$, where
$[\ldots,\ldots]_{\pm}$ is the commutator for $X= \tilde{B}$ and anticommutator for $X = c$ and $\overline{c}$, lead to
\bea
\left[ \varpi_{X(0)}(t), X_{(0)}(t)\right]_{\pm} &= & - \frac{i}{V},\\
\left[ \varpi_{X(+)}(t,\mathbf{x}), X_{(+)}(t,\mathbf{x}')\right]_{\pm} & = & -i\frac{N}{\sqrt{\gamma}}\delta(\mathbf{x},\mathbf{x}') + \frac{i}{V},\\
\left[  \varpi_{X(+)}(t,\mathbf{x}), X_{(0)}(t))\right]_{\pm} & = & \left[ \varpi_{X(0)}(t), X_{(+)}(t,\mathbf{x}')\right]_{\pm} = 0. \nonumber \\
\eea
Thus, the variables $V\varpi_{\tilde{B}(0)}$, $V\varpi_{c(0)}$ and $V\varpi_{\bar{c}(0)}$ are the canonical conjugate momenta of the zero modes
$\tilde{B}_{(0)}$, $c_{(0)}$ and $\bar{c}_{(0)}$, respectively.  Therefore,
in the functional Schr\"odinger representation, where states are represented as
functionals of $B(\mathbf{x})$, $c(\mathbf{x})$ and $\bar{c}(\mathbf{x})$, these operators are expressed as
\be
V\varpi_{X(0)} = - i \frac{\partial\ }{\partial X_{(0)}},\,\,\, X = \tilde{B}, c, \bar{c}.  \label{functional-Schroedinger}
\ee
Now, we have
\bea
V\varpi_{\tilde{B}(0)} &  = & - Q_A, \label{cons-eq1}\\
V\varpi_{c(0)} & = & iQ_{D\bar{c}}, \label{cons-eq2}\\
V\varpi_{\bar{c}(0)} & = &  -i\left[ Q_{Dc}  - q c_{(0)}\times Q_A\right]. \label{cons-eq3}
\eea
Thus, the conditions~\eqref{the-conditions} are equivalent to
\be
\varpi_{\tilde{B}(0)} |\textrm{phys}\rangle = \varpi_{c(0)}|\textrm{phys}\rangle = \varpi_{\bar{c}(0)}|\textrm{phys}\rangle = 0.
\label{the-condition2}
\ee
These conditions imply that a state represented as a wave functional of $\tilde{B}$, $c$ and $\overline{c}$ does not depend on the zero modes,
$\tilde{B}_{(0)}$, $c_{(0)}$ or $\bar{c}_{(0)}$. Thus, by requiring these
conditions for the physical states, we can eliminate the FP zero modes.
(It is necessary to redefine the inner product among the physical states.
This point is discussed in Appendix~\ref{appendix:group-averaging}.)

Next, we find the Hamiltonian obtained by eliminating the zero modes of $\tilde{B}$, $c$ and $\bar{c}$ in this manner.
The Hamiltonian written in terms of the field variables and their time derivatives (rather than their canonical conjugate momentum densities)
is invariant under the field redefinition~\eqref{field-redefinition} provided that the metric
is of the form~\eqref{our-metric} with the property~\eqref{gamma-factorization} as shown in Appendix~\ref{appendix:more-general-case}.
Thus, one can solve Eqs.~\eqref{varpi-c} and \eqref{varpi-bar-c} for $\nabla_0 c$ and
$\nabla_0 \overline{c}$ and substitute the resulting expressions
into Eq.~\eqref{H-GF+FP-before} to find the Hamiltonian.  Then, we find the effective Hamiltonian
$H_{\textrm{eff}}$ applicable to
the physical states satisfying the conditions~\eqref{the-condition2} by letting $\varpi_{X(0)} = 0$, i.e.\ $\varpi_{X} = \varpi_{X(+)}$,
$X=\tilde{B}, c, \overline{c}$.  The result is
\be
H_{\textrm{eff}}  = \int d\mathbf{x}\,\left(\mathcal{H}_{\textrm{class}} + \mathcal{H}_{\textrm{GF}+\textrm{FP}}\right)
+ H^{\prime}_{\textrm{GF}+\textrm{FP}},  \label{effective-H}
\ee
where
\bea
\mathcal{H}_{\textrm{GF}+\textrm{FP}}
& = & \frac{\sqrt{\gamma}}{N}\varpi_{c(+)}\cdot  (i\varpi_{\bar{c}(+)} + q\varpi_{\tilde{B}(+)} \times c_{(+)}) \nonumber \\
& &  - N\sqrt{\gamma}\,\gamma^{ij}\left(  i \nabla_i \overline{c}_{(+)}\cdot D_j c_{(+)} +  \nabla_i \tilde{B}_{(+)} \cdot A_j\right)  \nonumber \\
\eea
and
\be
H^{\prime}_{\textrm{GF}+\textrm{FP}} = - \frac{iq^2}{V}\overline{F}\cdot F, \label{extra-term}
\ee
with
\bea
 F & : = & \int d\mathbf{x}\,\frac{\sqrt{\gamma}}{N}( \varpi_{\tilde{B}(+)}\times \overline{c}_{(+)}),\\
\overline{F} & : = &   \int d\mathbf{x}\,\frac{\sqrt{\gamma}}{N} (\varpi_{\tilde{B}(+)}\times c_{(+)}).
\eea

The effective Lagrangian corresponding to this effective Hamiltonian is
\be
L_{\textrm{eff}} = \int d\mathbf{x}\,\mathcal{L}^{(+)}_{\textrm{Lan}} - H^{\prime}_{\textrm{GF}+\textrm{FP}},  \label{effective-Lagrangian}
\ee
where $\mathcal{L}^{(+)}_{\textrm{Lan}}$ is the Lagrangian density obtained by replacing $A_0$, $B$, $c$ and $\overline{c}$ by
$A_{0(+)}$, $\tilde{B}_{(+)}$, $c_{(+)}$ and $\overline{c}_{(+)}$, respectively.  Thus, we have an effective theory with the zero modes removed from these
fields and with an additional nonlocal interaction term $-H^{\prime}_{\textrm{GF}+\textrm{FP}}$ in the Lagrangian.

\section{Equivalence with the Faizal-Higuchi proposal}\label{sec:comparison}

In this section we describe the proposal by Faizal and Higuchi~\cite{Faizal2008} to solve the IR problem
caused by the FP ghosts in de~Sitter space
and show that their proposal is equivalent to the use of the effective Lagrangian~\eqref{effective-Lagrangian}.
They start from the FP-ghost propagator for a Lagrangian density with a small mass term $\sqrt{\left|g\right|}iM^2\overline{c}c$,
\be
T\langle 0| c^a(x)\overline{c}^b(x')|0\rangle = i\delta^{ab}D_M(x,x').
\ee
Here, the state $|0\rangle$ is the perturbative Bunch-Davies vacuum state and where
$D_M(x,x')$ is the propagator of the minimally coupled massive scalar field with mass $M$.  This propagator can be given in terms of the
variable $z = \cos^2(H\mu(x,x')/2)$, where $\mu(x,x')$ is the geodesic distance between the two points $x$ and $x'$ when they are spacelike separated.
(The function $\mu(x,x')$ can be analytically continued to the cases where $x$ and $x'$ are not connected by a spacelike geodesic.)
It is given by~\cite{AllenJacobson}
\be
D_M(x,x')  = \frac{H^{n-2}\Gamma(b_{0+})\Gamma(b_{0-})}{(4\pi)^{n/2}\Gamma(\frac{n}{2})} F\left(b_{0+},b_{0-};\frac{n}{2};z\right),
\label{M-propagator}
\ee
where $b_{0\pm}$ are obtained by letting $\ell=0$ in Eq.~\eqref{b-ell-pm}. (Faizal and Higuchi work only on four-dimensional de~Sitter space, but here
we generalize their proposal to $n$ dimensions with $n\geq 2$.)
For small $M$ we have
\be
D_M(x,x')  = \frac{1}{2c_0 M^2} + O(1),
\ee
where the constant $c_0$ is given by Eq.~\eqref{c0def}.

Now, the interaction term involving the FP ghosts in the Lagrangian density~\eqref{lagrangian3} is
$-i\sqrt{|g|}\nabla^\mu\overline{c}\cdot (A_\mu \times c)$, in which the FP antighost is differentiated.  This
means that, if we use the propagator~\eqref{M-propagator} in perturbation theory
and then take the massless limit, the divergent constant term of the propagator
does not contribute to the amplitude.  With this observation, Faizal and Higuchi proposed to use the effective propagator obtained by subtracting
the infinite constant from $D_M(x,x')$:
\bea
&& T\langle 0| c^a(x)\overline{c}^b(x')|0\rangle_{\textrm{eff}}\nonumber \\
&& = i\delta^{ab}
\lim_{M\to 0^+}\left[ D_M(x,x') - \frac{1}{2c_0 M^2}\right]. \label{effective-ghost-propagator}
\eea
Note here that this propagator is de~Sitter-invariant and IR finite.

We next describe this procedure of subtracting the IR-divergent constant
in terms of mode expansion of the propagator.  We discuss it  in any spacetime with the metric of the form~\eqref{our-metric} with $N=1$.
To start with we write down the Feynman propagator for the FP ghosts with $M\neq 0$ in terms of mode functions.
One first chooses a complete set of solutions,
$\varphi_{(n)}(t,\mathbf{x})$ (positive-frequency solutions) and $\varphi^*_{(n)}(t,\mathbf{x})$ (negative-frequency solutions), to the free-field equation
$(\nabla_\mu\nabla^\mu + M^2)\varphi_{(n)}(t,\mathbf{x}) = 0$   such that
\bea
i \int_{\Sigma}d\Sigma_\mu \left[ \varphi_{(n)}^*\nabla^\mu \varphi_{(m)}
- \varphi_{(m)}\nabla^\mu \varphi_{(n)}^*\right] & = & \delta_{mn}, \label{norm-cond}\\
i \int_{\Sigma}d\Sigma_\mu \left[ \varphi_{(n)}\nabla^\mu \varphi_{(m)} - \varphi_{(m)}\nabla^\mu \varphi_{(n)}\right] & = & 0.
\eea
Then the free FP-ghost field operators (with small mass) can be expanded as
\bea
c^a(t,\mathbf{x}) & = & \sum_{n} \left[ \alpha_n^a\varphi_{(n)}(t,\mathbf{x}) + \alpha_n^{a\dagger}\varphi_{(n)}^*(t,\mathbf{x})\right],\\
\overline{c}^a(t,\mathbf{x}) & = & \sum_{n} \left[ \overline{\alpha}_n^a\varphi_{(n)}(t,\mathbf{x}) + \overline{\alpha}_n^{a\dagger}
\varphi_{(n)}^*(t,\mathbf{x})\right].
\eea
The equal-time canonical anticommutation relations are
\bea
\left\{ \overline{c}^a(t,\mathbf{x}),\dot{c}^b(t,\mathbf{x}')\right\}
 & = & -\left\{\dot{\overline{c}}^a(t,\mathbf{x}),c^b(t,\mathbf{x}')\right\}\nonumber \\
& =  & \frac{\delta^{ab}}{\sqrt{\gamma(t,\mathbf{x})}}\delta(\mathbf{x},\mathbf{x}'),
\eea
with all other equal-time anticommutators of
the FP ghosts and their time derivatives vanishing. (Recall that we are assuming $N=1$.)
These lead to
\be
\left\{ \alpha^a_n, \overline{\alpha}_m^{b\dagger}\right\} = - \left\{ \alpha_n^{a\dagger}, \overline{\alpha}_m^b\right\} = i\delta^{ab}\delta_{mn},
\ee
with all other anticommutators among $\alpha_n^a$, $\overline{\alpha}_m^a$ and their Hermitian conjugates vanishing.
The perturbative vacuum state $|0\rangle$ is
annihilated by $\alpha_n^a$ and $\overline{\alpha}_m^a$.
The Feynman propagator is then
\bea
&& T\langle 0|c^a(t,\mathbf{x})\overline{c}^b(t',\mathbf{x}')|0\rangle \nonumber \\
&& = i \delta^{ab}\left[ \theta(t-t')\sum_{n}\varphi_{(n)}(t,\mathbf{x})\varphi^*_{(n)}(t',\mathbf{x}') \right. \nonumber \\
&& \ \ \ \ \left. + \theta(t'-t)\sum_n \varphi_{(n)}(t',\mathbf{x}')\varphi_{(n)}^*(t,\mathbf{x})\right],  \label{zero-mode-general}
\eea
where $\theta(t-t')$ is the Heaviside step function.

The Klein-Gordon equation for the zero mode reads
\be
\frac{1}{V(t)}\frac{d\ }{d t}\left[ V(t)\frac{d\varphi_{(0)}}{dt}\right]  + M^2\varphi_{(0)} = 0,
\ee
where $V(t)$ is given by Eq.~\eqref{vol-original} with $N=1$.  Two independent solutions with $M=0$ are $1$ and $f(t)$, which is defined
by Eq.~\eqref{ft}.  Hence, for small $M$  the solution $\varphi_{(0)}$ can be given as
\bea
\varphi_{(0)}(t) & \approx & \frac{1}{2C(M)}\left\{ 1 - M^2 [g(t)+c_3]\right\}\nonumber \\
&&  - i B(M)\left[ f(t)  + O(M^2)\right],  \label{phi_0}  \label{BMCM}
\eea
where the function $g(t)$ is given by Eq.~\eqref{gt}.  We choose $C(M)$ to be real and positive, but $B(M)$ may be complex.
In the $M\to 0^+$ limit the $O(M^2)$ contribution in the last term can be neglected relative to the term proportional to $g(t)+c_3$
as long as $B(M)C(M)\to 0$ as $M\to 0^+$, which we assume.
This is the case for the normalized positive-frequency zero modes in de~Sitter space, where $B(M)$ and $C(M)$ are both of order $M$ (see
Sec.~\ref{sec:IR-in-de-Sitter}), and on the static flat torus, where
they are of order $M^{1/2}$ (see Appendix~\ref{appendix:flat-torus}).

The normalization condition~\eqref{norm-cond} implies
\be
\lim_{M\to 0^+}\frac{\mathrm{Re}\,[B(M)]}{C(M)} = 1. \label{normalization-condition}
\ee
 We assume that $B(M)$ and $C(M)$ are of order $M^\alpha$ with $0 < \alpha \leq 1$. (For the de~Sitter case we have $\alpha=1$ whereas
for the static flat torus we have $\alpha=1/2$ as we pointed out above.)
Thus, $B(M)$, $C(M)$  and $M^2/C(M)$ tend to zero as $M\to 0^+$.  However, the limit  of $M/C(M)$ as $M\to  0^+$ is nonzero if $\alpha=1$.

The zero-mode sector of the Feynman propagator for the FP ghosts is IR divergent, and its divergent term reads
\bea
D^{(0)ab}(t,t') & = & i\delta^{ab}\left[\theta(t-t')\varphi_{(0)}(t)\varphi_{(0)}^*(t')  + (t\leftrightarrow t')\right] \nonumber \\
& = & \frac{i\delta^{ab}}{4C^2(M)} + O(1).  \label{zero-mode-M}
\eea
The subtraction of the IR-divergent term from the FP-ghost propagator changes the zero-mode contribution as follows:
\bea
D^{(0)ab}_{\textrm{eff}}(t,t') &: = &\lim_{M\to 0^+}\left[  D^{(0)ab}(t,t') - \frac{i\delta^{ab}}{4C^2(M)} \right]\nonumber \\
& = &  \delta^{ab}\left\{- i \beta_0^2 \left[g(t)+g(t')\right] + \frac{i \beta_1}{2}\left[f(t)+f(t')\right]\right. \nonumber \\
&&  \left. + \frac{1}{2}\left[\theta(t-t')-\theta(t'-t)\right]\left[f(t)-f(t')\right]\right\}, \nonumber \\
\label{effective-zero-mode}
\eea
where the constants $\beta_0,\,\beta_1$ are given by
\bea
\beta_0  & := &
\lim_{M\to 0^+} \frac{M}{C(M)},\\
\beta_1 &  := & \lim_{M\to 0^+}\frac{\mathrm{Im}\,[B(M)]}{C(M)}.
\eea
We have used Eq.~\eqref{normalization-condition} in the last term of Eq.~\eqref{effective-zero-mode}.
If $C(M)=O(M^\alpha)$ with $\alpha< 1$, then $\beta_0=0$.
 For the de~Sitter case we have $\alpha=1$ and
$\beta_0^2 = 2/c_0$ [see Eq.~\eqref{answerF}].

Now, the interaction term in the Lagrangian density involving the FP ghosts can be written as
$$
-i \sqrt{|g|}\,q\nabla_\mu \overline{c}\cdot (A^\mu \times c)\,\,\textrm{or}\,\, i\sqrt{|g|}\,q \nabla_\mu c \cdot (A^\mu \times \overline{c}).
$$
 Here, we have assumed
that one can freely integrate by parts at vertices with no boundary terms.  This might appear problematic in de~Sitter space, where the boundaries at the
past and future grow exponentially.  However, it turns out that one can construct an in-in formalism~\cite{Schwinger,Keldysh}
for which the boundary terms vanish upon integration by
parts in this spacetime~\cite{private}. (For early use of the in-in formalism in curved spacetime, see, e.g.\ Refs.~\cite{Hajicek,Kay-in-in,Jordan,CalzettaHu}.)
 Therefore, we proceed under the assumption that integration by parts  does not generate nonzero surface terms at the vertex  where the FP ghosts interact
with the gauge field.   (We emphasize, however,
that this assumption is necessary only for the equivalence of our method and the Faizal-Higuchi proposal.   Our method itself is valid
even if integration by parts
generates nonzero surface terms.)  Note also that we have set $\nabla_\mu A^\mu = 0$.  This is valid because in the Landau
gauge the Feynman propagator for the gauge field is divergence-free.

The interaction term involving the zero modes of the FP ghosts can then be written as
\bea
I_c & = & -i \sqrt{|g|}\,q\partial_0 \overline{c}_{(0)}\cdot (\varpi_{\tilde{B}(+)}\times c_{(+)}),\label{int1}\\
I_{\bar{c}} & = & - i\sqrt{|g|}\, q(\varpi_{\tilde{B}(+)}\times \overline{c}_{(+)})\cdot \partial_0 c_{(0)}. \label{int2}
\eea
We have dropped the terms that become zero upon integration over the spacetime, or more precisely, over the
space and the chosen time path in the appropriate in-in formalism.  For example, terms
of the form $X_{(0)}Y_{(0)}Z_{(+)}$ vanish upon integration over the space.
(In general we need to subtract the zero-mode contribution to $A_0$ from the propagator of the gauge field to impose the condition
$A_{0(0)}|0\rangle = - \varpi_{\tilde{B}(0)}|0\rangle = 0$ on the perturbative vacuum $|0\rangle$.
Interestingly, for the de~Sitter case this condition follows automatically from de~Sitter invariance  as shown in
Appendix~\ref{appendix:zero-mode-for-A}.)
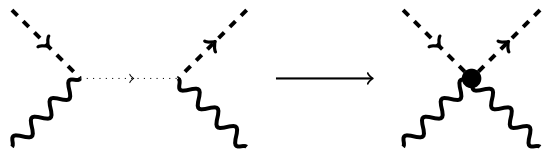
\begin{figure}
\begin{center}
\begin{tikzpicture}[line width=1.5 pt, scale=1.3]
	\draw[vector] (0.7,-0.7)--(0,0);
	\draw[scalarbar] (0.7,0.7)--(0,0);
\draw[scalar] (-1.7,0.7) -- (-1,0);
	\draw[zeroscalar,thin] (180:1)--(0,0);
\draw[vector] (-1.7,-0.7) -- (-1,0);
\draw[thick,->] (1,0) -- (2,0);
\draw[vector] (3.7,-0.7)--(3,0);
	\draw[scalarbar,] (3.7,0.7)--(3,0);
\draw[scalar] (2.3,0.7) -- (3,0);
\draw[vector] (2.3,-0.7) -- (3,0);
\fill (3,0) circle (1mm);
 \end{tikzpicture}
\caption{The wavy, dashed and dotted lines represent the gauge field, the
nonzero-mode part of the FP-ghost propagator and its zero-mode part, respectively. The zero-mode contribution to the
FP-ghost propagator
is integrated out in Eq.~\eqref{almost-end} by using Eq.~\eqref{delta-result}.}
 \end{center}
\end{figure}

Since the zero-mode contribution to the fields $c$ and $\overline{c}$ appears only once in each of the interaction terms~\eqref{int1} and \eqref{int2},
one can integrate out the zero-mode contribution in perturbation theory as shown in Fig.~1.
This integration introduces the following extra term
in the Lagrangian:
\bea
L^{\textrm{extra}}(t)  & = & -i \int d\mathbf{x}\,\sqrt{|\gamma(t,\mathbf{x})|}\int dt' d\mathbf{x}'\,\sqrt{|\gamma(t',\mathbf{x})|}\nonumber \\
&& \times \left[\varpi_{\tilde{B}(+)}(t,\mathbf{x})\times \overline{c}_{(+)}(t,\mathbf{x})\right]^a
\frac{\partial^2\ }{\partial t\partial t'}D^{(0)ab}_{\textrm{eff}}(t,t')\nonumber \\
&&  \times
\left[ \varpi_{\tilde{B}(+)}(t',\mathbf{x}')\times c_{(+)}(t',\mathbf{x}')\right]^b. \label{almost-end}
\eea
Now we find from Eq.~\eqref{effective-zero-mode}
\be
\frac{\partial^2\ }{\partial t\partial t'}D^{(0)ab}_{\textrm{eff}}(t,t') = - \delta^{ab}\frac{1}{V(t)}\delta(t-t').  \label{delta-result}
\ee
By substituting this expression into Eq.~\eqref{almost-end} we obtain
$L^{\textrm{extra}} = - H^{\prime}_{\textrm{GF}+\textrm{FP}}$, where $H^{\prime}_{\textrm{GF}+\textrm{FP}}$ is given by
Eq.~\eqref{extra-term}.  Thus, Lagrangian perturbation theory
with the effective FP-ghost propagator~\eqref{effective-ghost-propagator} proposed in Ref.~\cite{Faizal2008} agrees with perturbation theory
using the effective Lagrangian~\eqref{effective-Lagrangian}, which has been derived by imposing the conditions~\eqref{the-conditions}.

\section{Summary and discussion}\label{sec:summary}

In this paper we showed that the space-independent modes
of the Faddeev-Popov ghosts can be removed from the covariantly quantized Yang-Mills theory in spacetime
with compact spatial sections in the Landau gauge by requiring that the Noether charges corresponding to the conserved currents $A^\mu$, $D^\mu c$ and
$D^\mu \overline{c}$ annihilate the physical states.  There will be no IR divergences or
de~Sitter symmetry breaking from the ghost sector if these conditions are applied to
Yang-Mills theory in global de~Sitter space.  We also showed that the compatibility of these conditions with the invariance of the physical states under
the BRST transformation leads to the requirement that the physical states be invariant under the global transformations corresponding to the
gauge group.  Then we showed that the theory resulting from removing the FP-ghost zero modes in this manner is equivalent to using the effective FP-ghost
propagator with the constant IR-divergent term subtracted.  This propagator is de~Sitter-invariant, which confirms that our method of removing the zero modes
is de~Sitter-invariant when applied to de~Sitter space.

The IR issues in perturbative gravity in de~Sitter space have been debated since the physical~\footnote{Here, the two-point function
is called ``physical'' in the sense that the gauge degrees of freedom are completely fixed.}
graviton two-point function in the conformally flat, or Poincar\'e, patch was found to be IR
divergent~\cite{FordParker1977b}.
However, it has been demonstrated that
the IR divergences in this two-point function are gauge artifact in the sense that they can be expressed in
a pure-gauge form~\cite{HiguchiNP,AllenNP,HiguchiKouris2000}
and that the mode functions can be chosen to make the two-point function
IR finite~\cite{HiguchiMarolfMorrison}.  IR-finite physical graviton two-point functions have also
been constructed in other coordinate patches of de~Sitter space~\cite{HawkingHertogTurok,HiguchiWeeks,BernarCrispinoHiguchi}.  The IR-finite graviton
propagators in covariant gauges have also been derived~\cite{AllenTuryn,HiguchiKouris2001,Morrison2013}.
It has been shown that these propagators can be constructed by the mode-sum method
in the global patch of de~Sitter space provided that the Euclidean \emph{massive} tensor-field propagator is obtained using this method~\cite{Faizal2012}.
(The mode summation in the Poincar\'e patch, on the other hand, appears to lead to an IR-divergent graviton propagator even in the covariant
gauge~\cite{TsamisWoodard1,TsamisWoodard2} though the IR-divergent terms are of pure-gauge form.)
Moreover, the gauge-invariant two-point function as defined in Ref.~\cite{FewsterHunt} is known to be equivalent to that of the
linearized Weyl tensor~\cite{Higuchi-unpublished}, which is known to be IR finite and de~Sitter-invariant in the
Bunch-Davies-like vacuum state~\cite{Kouris,MoraTsamisWoodard,Froeb}.
Thus, in fact there are no IR problems in \emph{noninteracting} linearized gravity.
However, it is not clear whether or not the IR divergences of the graviton propagator in the Poincar\'e patch will lead to physical effects, such as the breakdown
of de~Sitter symmetry, in perturbative gravity.  Some authors claim they will
(see, e.g.\ Ref.~\cite{TsamisWoodardbreaking}) but other authors suggest they will not (see, e.g.\ Ref.~\cite{TanakaUrakawa}).

We note in this context that the de~Sitter-invariant FP-ghost propagator in the covariant gauge is also
IR divergent.  Hence, one might expect that the IR effect
in the FP-ghost sector would lead to breakdown of de~Sitter symmetry.  However,
the interaction terms involving the FP ghosts are
such that  the IR-divergent part of the propagator does not contribute if one regularizes the IR divergences by a small mass term
as in the Yang-Mills case treated in this paper.  Because of this fact
it was proposed in Ref.~\cite{Faizal2008} that one should simply subtract the IR-divergent contribution from the FP-ghost
propagator to obtain an effective de~Sitter
invariant propagator, as in the Yang-Mills case.

Now, it is not difficult to see that there are conserved Noether charges analogous to those found in Sec.~\ref{sec:conserved-charges}
in covariantly quantized perturbative gravity in the Landau gauge.  Thus, we expect that the modes of the FP ghosts responsible for IR divergences
will be removed if the physical states are required to be annihilated by these conserved charges.  If this is the case,
it will be interesting to find out whether or not the effective
theory thus obtained is equivalent to the proposal of Ref.~\cite{Faizal2008} in perturbative gravity as in Yang-Mills theory.  This question is currently
under investigation.

\begin{acknowledgments}
We thank Mir Faizal, Chris Fewster, Markus Fr\"ob and Albert Roura for useful discussions.  The work of J.\ G. was
supported by the Engineering and Physical Sciences Research Council (EPSRC).
\end{acknowledgments}

\appendix
\section{The zero mode of the minimally coupled scalar field in de~Sitter space}\label{appendix:deSitterzeromode}

Let us first examine the small $M$ limit of the hypergeometric function in Eq.~\eqref{secondary} with $\ell=0$.  We use the
formula 9.136.1 in Ref.~\cite{GradshteynRyzhik},
\bea
&& F\left( 2\alpha,2\beta;\alpha+\beta+\frac{1}{2}; \frac{1-\sqrt{z}}{2}\right) \nonumber \\
&& = AF\left(\alpha,\beta;\frac{1}{2};z\right) + B\sqrt{z}F\left(\alpha+\frac{1}{2},\beta+\frac{1}{2};\frac{3}{2};z\right), \nonumber \\
\eea
where
\bea
A & = & \frac{\Gamma(\alpha+\beta+\frac{1}{2})\sqrt{\pi}}{\Gamma(\alpha+\frac{1}{2})\Gamma(\beta+\frac{1}{2})}, \\
B & = & - \frac{\Gamma(\alpha+\beta+\frac{1}{2})2\sqrt{\pi}}{\Gamma(\alpha)\Gamma(\beta)},
\eea
with $z= -\sinh^2 t$ and $\sqrt{z} = i\sinh t$.  We find
\bea
&& F\left( b_{0+},b_{0-},\ell+\frac{n}{2};\frac{1-i\sinh Ht}{2}\right) \nonumber \\
&&  = A F\left( \frac{b_{0+}}{2},\frac{b_{0-}}{2};\frac{1}{2}; - \sinh^2 Ht\right) \nonumber \\
&& \,\, + i B\sinh Ht\,F\left( \frac{n}{2},\frac{1}{2};\frac{3}{2}; - \sinh^2 Ht\right), \label{hyper-eq}
\eea
where
\bea
A & \approx &  1 + \left[\psi(\tfrac{n}{2}) - \psi(\tfrac{1}{2})\right]\frac{M^2}{2(n-1)H^2}, \label{Aapprox}\\
B & \approx & - \frac{\Gamma(\frac{n}{2})\sqrt{\pi}M^2}{2\Gamma(\frac{n+1}{2})H^2}. \label{Bapprox}
\eea
Recall that $\psi(x):= \Gamma'(x)/\Gamma(x)$ and
\bea
b_{0\pm} & = & \frac{n-1}{2} \pm \sqrt{\left(\frac{n-1}{2}\right)^2 - \frac{M^2}{H^2}} \nonumber \\
& \approx & \frac{n-1}{2} \pm \left( \frac{n-1}{2} - \frac{M^2}{(n-1)H^2}\right).
\eea
One can readily show that
\bea
&& \frac{1}{H}\frac{d\ }{dt}
\left[ \sinh Ht\, F\left( \frac{n}{2},\frac{1}{2};\frac{3}{2};-\sinh^2 Ht\right)\right],\nonumber \\
&& = (\cosh H t)^{-(n-1)}
\eea
using the series expression of the hypergeometric function.  Thus,
\bea
&& \sinh Ht\, F\left(\frac{n}{2},\frac{1}{2};\frac{3}{2}; -\sinh^2 Ht\right)\nonumber \\
&& = H\int_0^t \frac{dt'}{(\cosh Ht)^{n-1}}. \label{first-hyper}
\eea
We find in a similar manner, noting that $b_{0-}|_{M=0} = 0$ and using Eq.~\eqref{elementary},
\bea
&& F\left( \frac{b_{0+}}{2},\frac{b_{0-}}{2}; \frac{1}{2}; - \sinh^2 Ht\right) \nonumber \\
&& \approx 1 -  M^2 \int_0^t dt'\,  (\cosh Ht')^{-(n-1)}\nonumber \\
&& \ \ \ \ \ \ \ \ \ \ \ \ \ \ \ \ \ \ \times \int_0^{t'} dt'' (\cosh Ht'')^{n-1}.  \label{second-hyper}
\eea
Then, substituting Eqs.~\eqref{Aapprox}, \eqref{Bapprox}, \eqref{first-hyper} and \eqref{second-hyper} into Eq.~\eqref{hyper-eq}, we find
\bea
&& F\left( b_{0+},b_{0-};\ell+\frac{n}{2};\frac{1-i\sinh Ht}{2}\right)\nonumber \\
&& \approx  1  - M^2\left\{ g(t) - \left[\psi(\tfrac{n}{2}) - \psi(\tfrac{1}{2})\right]\frac{1}{2(n-1)H^2}\right\} \nonumber \\
&& \,\,\,\, - ic_0 M^2 f(t), \label{F-Mzero}
\eea
where the functions $f(t)$ and $g(t)$ are defined by Eqs.~\eqref{ft} and \eqref{gt}, respectively, and the constant $c_0$ is given by
Eq.~\eqref{c0def}.

The normalized zero mode $F_0(t) = f_0(t)Y_{00}(\mbox{\boldmath$\theta$})$, where $f_0(t)$ can be found by letting $\ell=0$ in
Eq.~\eqref{secondary}, is obtained by multiplying the hypergeometric function in Eq.~\eqref{F-Mzero} by
\be
K_0 =  \frac{H^{\frac{n-2}{2}}N_0}{2^{\frac{n-2}{2}} \Gamma(\frac{n}{2})}Y_{00}(\mbox{\boldmath$\theta$}).
\ee
Recalling that
$Y_{00}(\mbox{\boldmath$\theta$}) = V_{S^{n-1}}^{-1/2}$, where $V_{S^{n-1}}$ is the volume of the unit $S^{n-1}$ given by
Eq.~\eqref{V-unit-sphere},
and using the expression of $N_\ell$ with $\ell=0$ in Eq.~\eqref{Nell} (choosing the convention $N_0>0$), we find
\be
K_0 = \left(\frac{H^{n-2} \Gamma(\frac{n-1}{2}+\lambda)\Gamma(\frac{n-1}{2}-\lambda)}{2^n \pi^{n/2}\Gamma(\frac{n}{2})}\right)^{1/2}.
\ee
For small $M$ we obtain
\be
K_0  \approx \frac{1}{\sqrt{2c_0}M}\left\{ 1 - \left[ \psi(n-1)-\psi(1)\right]\frac{M^2}{2(n-1)H^2}\right\}, \label{K-0-approx}
\ee
where we have used the doubling formula:
\be
\Gamma(2x) = \frac{2^{2x-1}}{\sqrt{\pi}}\Gamma(x)\Gamma(x+\tfrac{1}{2}). \label{Gamma2x}
\ee
Multiplying Eq.~\eqref{F-Mzero} by $K_0$ in Eq.~\eqref{K-0-approx}, we indeed obtain Eq.~(\ref{answerF}) with
\be
c_1 = \frac{1}{2(n-1)H^2}\left[ \psi(n-1) - \psi(\tfrac{n}{2}) - \psi(1) + \psi(\tfrac{1}{2})\right]. \label{c-1-value}
\ee

Next we examine the action of the boost Killing vector $L_X$ defined by Eq.~\eqref{LX-def} on the mode functions
$F_{\ell m \tilde{\sigma}}(t,\mbox{\boldmath$\theta$})$ and derive Eq.~\eqref{LX-eq},
following Ref.~\cite{HiguchiJMP}.
We first note that the Killing vector $L_X$ can be written as
\be
L_X = T^{(+)}S^{(+)}  +  T^{(-)}S^{(-)},  \label{LX-decompose}
\ee
where
\bea
T^{(+)} & = & \frac{1}{H}\frac{\partial\ }{\partial t} - \ell \tanh Ht, \\
T^{(-)} & = & \frac{1}{H} \frac{\partial\ }{\partial t} +  (\ell + n - 2)\tanh Ht,
\eea
and
\bea
S^{(+)} & = & \frac{1}{2\ell + n -2}\left[ \sin\chi \frac{\partial\ }{\partial\chi} + (\ell+n-2)\cos\chi\right], \nonumber \\ \\
S^{(-)} & = & - \frac{1}{2\ell + n -2}\left[ \sin\chi \frac{\partial\ }{\partial\chi} - \ell \cos\chi\right].
\eea
By using the raising and lowering operators for the associated Legendre functions found in 8.733.1 of Ref.~\cite{GradshteynRyzhik}, we obtain
\bea
T^{(+)}f_\ell(t) & = & - i \left[\ell(\ell+n-1) + \frac{M^2}{H^2}\right]^{1/2}f_{\ell + 1}(t), \\
T^{(-)}f_\ell(t) &  = & - i \left[ (\ell-1)(\ell+n-2) + \frac{M^2}{H^2}\right]^{1/2} f_{\ell-1}(t),\nonumber \\
\label{Tminus}
\eea
where Eq.~\eqref{Tminus} is for $\ell \geq 1$, and
\bea
S^{(+)}Y_{\ell m \tilde{\sigma}} & = & \left[ \frac{(\ell-m+1)(\ell+m+n-2)}{(2\ell+n)(2\ell+n-2)}\right]^{1/2}Y_{(\ell+1) m \tilde{\sigma}},\nonumber \\ \\
S^{(-)}Y_{\ell m \tilde{\sigma}} & = & \left[ \frac{(\ell -m)(\ell+m+n-3)}{(2\ell+n-2)(2\ell+n-4)}\right]^{1/2} Y_{(\ell-1)m \tilde{\sigma}}. \nonumber \\
\eea
By using these formulas and the decomposition~\eqref{LX-decompose} we find Eq.~\eqref{LX-eq}.

\section{The effective Hamiltonian in general spacetime}\label{appendix:more-general-case}

In this appendix we present the effective Hamiltonian in spacetime with the more general line element of the Arnowitt-Deser-Misner metric~\cite{Witten}, given by
\be
ds^2 = N^2dt^2 - \gamma_{ij}(dx^i + N^i dt)(dx^j + N^j dt),
\ee
where the lapse function $N$, the shift vector $N^i$ and the metric $\gamma_{ij}$ on space
depend on $t$ and $\mathbf{x}$ in general.  The standard Hamiltonian density $\mathcal{H}_{\textrm{st}}$
is given as
\bea
\frac{1}{\sqrt{|g|}}\mathcal{H}_{\textrm{st}}
&  = & \frac{1}{\sqrt{|g|}}\mathcal{H}_{\textrm{class}} + A^i \cdot \left(\nabla_i B - iq\nabla_i\overline{c}\times c\right) \nonumber \\
&& - \frac{i}{N^2}\nabla_0\overline{c}\cdot\nabla_0 c \nonumber \\
&& - i\left( \gamma^{ij} - \frac{N^iN^j}{N^2}\right) \nabla_i \overline{c}\cdot\nabla_j c, \label{standard-H}
\eea
where the classical Hamiltonian density $\mathcal{H}_{\textrm{class}}$ is given by the following formula:
\be
\frac{1}{\sqrt{|g|}}\mathcal{H}_{\textrm{class}}
= - \frac{1}{2}F_{0i}\cdot F^{0i} + \frac{1}{4}F_{ij}\cdot F^{ij} + \frac{1}{N^2}D_i A_0 \cdot \pi^i,
\ee
with
\be
 F^{0i} = -N^{-2}\pi^i.
\ee
We again define $\tilde{B}:= B - iq \overline{c}\times c_{(0)}$,
where $c_{(0)}$ is defined by Eq.~\eqref{define-c0} with $V(t)$ given by Eq.~\eqref{vol-original}.
We again define the conjugate momentum densities of $\tilde{B}$, $c$ and $\overline{c}$ multiplied by $N/\sqrt{\gamma}$ as $\varpi_{\tilde{B}}$,
$\varpi_c$ and $\varpi_{\bar{c}}$. They are given by
\bea
\varpi_{\tilde{B}} & = & - N^2 A^0, \label{varpi1}\\
\varpi_c & = & i N^2 \nabla^0\overline{c} - \frac{iq}{V(t)}\int d\mathbf{x}\,\frac{\sqrt{\gamma}}{N}(\varpi_{\tilde{B}}\times \overline{c}),\label{varpi2}\\
\varpi_{\bar{c}} & = & -i N^2\left(\nabla^0 c + q A^0 \times c_{(+)}\right). \label{varpi3}
\eea
The zero modes $\varpi_{\tilde{B}(0)}$, $\varpi_{c(0)}$ and $\varpi_{\bar{c}(0)}$ defined as in
Sec.~\ref{sec:conserved-charges}
are related to the conserved charges $Q_A$, $Q_{Dc}$ and $Q_{D\bar{c}}$ as in Eqs.~\eqref{cons-eq1}-\eqref{cons-eq3}.  Therefore, we can
require the conditions $\varpi_{\tilde{B}(0)}|\textrm{phys}\rangle = \varpi_{c(0)}|\textrm{phys}\rangle = \varpi_{\bar{c}(0)}|\textrm{phys}\rangle = 0$
for the physical states $|\textrm{phys}\rangle$.

We have $\varpi_{\tilde{B}} = \varpi_{B}$ but the conjugate momentum densities for $c$ and
$\overline{c}$ differ from the ones before the variable change by the following quantities:
\bea
\Delta \varpi_c & = & - \frac{iq}{V(t)}\int d\mathbf{x}\,\frac{\sqrt{\gamma}}{N}(\varpi_{\tilde{B}}\times \overline{c}),\\
\Delta \varpi_{\bar{c}} & = & - iq\varpi_{\tilde{B}}\times c_{(0)}.
\eea
The Hamiltonian after the change of variables $\tilde{B} = B - iq\overline{c}\times c_{(0)}$ is different
from the standard Hamiltonian even when it is expressed in terms of the fields and their time derivatives (rather than
their canonical conjugate momenta)
because this change of variables explicitly depends on $t$.  We note that this is not the case with the metric chosen
in Sec.~\ref{sec:effective-Hamiltonian} because the definition of $c_{(0)}$ there does not depend on $t$ explicitly.
The difference $\Delta \mathcal{H}$
between the Hamiltonian density $\mathcal{H}$ after the variable change and the standard Hamiltonian density $\mathcal{H}_{\textrm{st}}$ is given as
\bea
\frac{1}{\sqrt{|g|}}\Delta \mathcal{H} & = & \frac{1}{\sqrt{|g|}}(\mathcal{H} - \mathcal{H}_{\textrm{st}}) \nonumber \\
& = & N^{-2} \frac{\partial\ }{\partial t}(-iq\overline{c}\times c_{(0)}) \cdot \varpi_{\tilde{B}} \nonumber \\
&& + N^{-2}( - \Delta \varpi_{c}\cdot \dot{c} + \dot{\overline{c}}\cdot \Delta \varpi_{\bar{c}}).
\eea
Then
\bea
\Delta H & := & \int d\mathbf{x}\,\Delta \mathcal{H} \nonumber \\
& = & - iq \int d\mathbf{x}\frac{\sqrt{\gamma}}{N}(\varpi_{\tilde{B}}\times \overline{c}) \nonumber \\
&& \cdot \left[\frac{\partial\ }{\partial t}\left(\int d\mathbf{y}\frac{\sqrt{\gamma}}{V(t)N} c(t,\mathbf{y})\right)
- \int d\mathbf{y}\frac{\sqrt{\gamma}}{V(t)N}\frac{\partial\ }{\partial t}c(t,\mathbf{y})\right] \nonumber \\
& = & - iq \int d\mathbf{x}\frac{\sqrt{\gamma}}{N}(\varpi_{\tilde{B}}\times \overline{c}) \cdot
\int d\mathbf{y}\frac{\partial\ }{\partial t}\left[ \frac{\sqrt{\gamma}}{V(t)N}\right]c_{(+)}(t,\mathbf{y}). \nonumber \\
\label{Delta-H-result}
\eea
It is clear that $\Delta H = 0$ for the metric chosen in Sec.~\ref{sec:effective-Hamiltonian} because $\sqrt{\gamma}/(VN)$ is time independent there.
We have replaced $c(t,\mathbf{y})$ by $c_{(+)}(t,\mathbf{y})$ because
\be
\int d\mathbf{y}\frac{\partial\ }{\partial t}\left[ \frac{\sqrt{\gamma}}{V(t)N}\right] = 0.
\ee

The effective Hamiltonian for the physical states $|\mathrm{phys}\rangle$ satisfying
$\varpi_{\tilde{B}(0)}|\mathrm{phys}\rangle = \varpi_{c(0)}|\mathrm{phys}\rangle = \varpi_{\bar{c}(0)}|\mathrm{phys}\rangle = 0$ can be obtained
by solving Eqs.~\eqref{varpi1}-\eqref{varpi3} for $A^0$, $\nabla_0\overline{c}$ and $\nabla_0 c$, and substituting the results and
$B = \tilde{B} + iq\overline{c}\times c_{(0)}$ in Eq.~\eqref{standard-H} and adding $\Delta H$ given by Eq.~\eqref{Delta-H-result}.
The result can be given in the following form:
\be
H_{\textrm{eff}} = \int d\mathbf{x}\,\mathcal{H}_{\textrm{conv}} - \frac{iq}{V(t)}\overline{F}\cdot \widetilde{F}.
\ee
Here, the "conventional" Hamiltonian density $\mathcal{H}_{\textrm{conv}}$ is given by
\bea
 \mathcal{H}_{\textrm{conv}}
&  = &   \mathcal{H}_{\textrm{class}} +
\sqrt{\gamma}N A^i \cdot (\nabla_i\tilde{B} - iq\nabla_i \overline{c}\times c)\nonumber \\
&& - \frac{\sqrt{\gamma}}{N}\varpi_{c}\cdot \left[ i\varpi_{\bar{c}} + N^i \nabla_i c + q\varpi_{\tilde{B}} \times c\right] \nonumber \\
&&  + \frac{\sqrt{\gamma}}{N}N^i \nabla_i \overline{c}\cdot (\varpi_{\bar{c}} - i q\varpi_{\tilde{B}}\times c) \nonumber \\
&&  + \sqrt{\gamma}N \gamma^{ij}\nabla_i \overline{c}\cdot \nabla_j c,
\eea
where the fields $c$, $\overline{c}$, $B$ and their (rescaled) conjugate momentum densities
$\varpi_c$, $\varpi_{\bar{c}}$ and $\varpi_{\tilde{B}}$ are understood to be their nonzero-mode contributions,
$c_{(+)}$, $\overline{c}_{(+)}$, $\tilde{B}_{(+)}$, $\varpi_{c(+)}$,
$\varpi_{\bar{c}(+)}$ and $\varpi_{\tilde{B}(+)}$.  The operators $\overline{F}$ and $\widetilde{F}$
are given by
\bea
\overline{F} & : = & \int d\mathbf{x}\, \frac{\sqrt{\gamma}}{N}(\varpi_{B(+)}\times \overline{c}_{(+)}),\\
\widetilde{F} & :  = & \int d\mathbf{x}\,\frac{\sqrt{\gamma}}{N}\nonumber \\
&& \times \left[ \frac{N}{\sqrt{\gamma}} \frac{\partial\ }{\partial t}\left( \frac{\sqrt{\gamma}}{N}\right)c_{(+)}
+ q\varpi_{\tilde{B}(+)}\times c_{(+)} + N^i \nabla_i c_{(+)}\right]. \nonumber \\
\eea
In principle one can use the Hamiltonian $H_{\textrm{eff}}$ here for perturbative calculations though it may not be very practical to do so.

\section{The Dirac bracket}\label{appendix:dirac-bracket}

In this paper we identified $-\sqrt{|g|}A^0$ with the canonical momentum density for $B$ (and $\tilde{B}$), and did not regard it as an independent
dynamical variable.
In this appendix we show that this treatment agrees with the quantization using the Dirac bracket. (The variables $B$ and $\pi_B$ may be replaced by
$\tilde{B}$ and $\pi_{\tilde{B}}$ below without any other changes.)  We suppress the argument $t$ since all brackets are computed at an equal time.

We have two second-class constraints :
\bea
C_1 & : = & |g|^{-1/2}\pi_B +  A^0 \approx 0,\\
C_2 &: = & \pi_{A^0} \approx 0,
\eea
where $\pi_{A^0}$ is the canonical conjugate momentum
density for $A^0$.  The Poisson bracket for these constraints is
\be
\{ C_I(\mathbf{x}),C_J(\mathbf{x}')\}_{\textrm{PB}} = \delta(\mathbf{x},\mathbf{x}')\epsilon_{IJ},
\ee
where $\epsilon_{IJ}$ is the antisymmetric $2\times 2$ matrix with $\epsilon_{12} = 1$.  The inverse of this Poisson bracket is
\be
\Delta^{IJ}(\mathbf{x},\mathbf{x'}) = - \delta(\mathbf{x},\mathbf{x}')\epsilon^{IJ},
\ee
where $\epsilon^{IJ}$ is identical with $\epsilon_{IJ}$ as a matrix.

The Dirac bracket for two canonical variables $X(\mathbf{x})$ and $Y(\mathbf{x}')$
can be computed as
\bea
&& \{ X(\mathbf{x}), Y(\mathbf{x}')\}_{\textrm{DB}}\nonumber \\
& & = \{X(\mathbf{x}),Y(\mathbf{x}')\}_{\textrm{PB}}\nonumber \\
&&  - \int d\mathbf{y}\int d\mathbf{z}\{X(\mathbf{x}),C_I(\mathbf{y})\}_{\textrm{PB}}
\Delta^{IJ}(\mathbf{y},\mathbf{z})
\{ C_J(\mathbf{z}),Y(\mathbf{x}')\}_{\textrm{PB}} \nonumber \\
&& =  \{X(\mathbf{x}),Y(\mathbf{x}')\}_{\textrm{PB}} \nonumber \\
&&  + \int d\mathbf{y}\{X(\mathbf{x}),C_I(\mathbf{y})\}_{\textrm{PB}}
\{ C_J(\mathbf{y}),Y(\mathbf{x}')\}_{\textrm{PB}}\epsilon^{IJ}.
\eea
Then, by construction $\{C_I(\mathbf{x}),X(\mathbf{x}')\}_{\textrm{DB}} = 0$ for all $C_I(\mathbf{x})$ and $X(\mathbf{x}')$.  We also have
\be
\{B(\mathbf{x}),\pi_B(\mathbf{x}')\}_{\textrm{DB}} = \{ B(\mathbf{x}),\pi_{B}(\mathbf{x}')\}_{\textrm{PB}} = \delta(\mathbf{x},\mathbf{x}').
\ee
Hence, if we "promote" the Dirac bracket to commutation relations, we find that the operators $\pi_{A^0}$ and $\pi_{B} + \sqrt{|g|}A^0$
commute with all operators and that $B$ and $\pi_B$ satisfy the standard commutation relations. Hence,  basing
the canonical commutation relations on the Dirac bracket is equivalent to letting $\pi_{A^0} = 0$ and $\pi_{B} = -\sqrt{|g|}A^0$, as we did in this paper.

\section{Redefining the inner product}\label{appendix:group-averaging}

As was pointed out in Sec.~\ref{sec:effective-Hamiltonian}, if we represent the state $|\textrm{phys}\rangle$ as a wave functional
$\Psi [\tilde{B}(\mathbf{x}),c(\mathbf{x}),\overline{c}(\mathbf{x})]$, then the conditions
$\varpi_{\tilde{B}(0)}|\textrm{phys}\rangle = \varpi_{c(0)}|\textrm{phys}\rangle = \varpi_{\bar{c}(0)}|\textrm{phys}\rangle =0$ can be translated
to the following equations:
\be
\frac{\partial\ }{\partial \tilde{B}_{(0)}}\Psi = 0 \label{bosonic-condition}
\ee
and
\be
\frac{\partial\ }{\partial c_{(0)}}\Psi = \frac{\partial\ }{\partial \bar{c}_{(0)}}\Psi = 0.  \label{fermionic-condition}
\ee

Now the wave functional $\Psi$ is formally normalized by the condition
\be
\int [d\tilde{B}][dc][d\overline{c}]|\Psi[\tilde{B},c,\overline{c}]|^2 = 1. \label{norm-int}
\ee
This functional integral contains the integrations over the variables $\tilde{B}_{(0)}$, $c_{(0)}$ and $\overline{c}_{(0)}$.  Since the functional $\Psi$ is
independent of these variables due to Eqs.~\eqref{bosonic-condition} and \eqref{fermionic-condition}, we have
\bea
\int d\tilde{B}_{(0)} |\Psi|^2  & = & \infty,\\
\int dc_{(0)} |\Psi|^2 & = & \int d\,\overline{c}_{(0)}|\Psi|^2 = 0. \label{fermion=0}
\eea
Equation~\eqref{fermion=0} follows from the rule that the integration with respect to a Grassmann variable is the same as the
(left) differentiation~\cite{Berezin,PeskinSchroeder}.
Thus, the normalization integral~\eqref{norm-int} will be (even more) ill-defined.  However,
this situation can readily be remedied by dropping the variables $\tilde{B}_{(0)}$, $c_{(0)}$ and $\overline{c}_{(0)}$ from the normalization integral.

In the rest of this appendix, we present a $2$-dimensional representation of the fermionic commutation relations
$\theta^2 = \pi_\theta^2 = 0$, $\{\pi_\theta,\theta \} = -i$, where $\theta^\dagger = \theta$.  We expect from Eq.~\eqref{fermion=0} that a state annihilated
by $\pi_\theta$ has zero norm.  It is clear that the nilpotency of the Hermitian
operator $\theta$ requires an indefinite pseudoinner product because the square of a nonzero
Hermitian operator cannot vanish in a Hilbert space with
a positive-definite inner product.
We introduce a pseudoinner product on the $2$-dimensional vector space $\textrm{Span}\{ |0\rangle,|1\rangle\}$ by
$\langle 0|0\rangle= \langle 1|1\rangle = 0$, $\langle 0|1\rangle = \langle 1|0\rangle = 1$.  We let
$\theta|0\rangle = |1\rangle$ and $\theta|1\rangle = 0$.  We find $\theta^2 = 0$ and that $\theta$ is Hermitian:
$\langle 0|\theta|1\rangle = \langle 1|\theta|0 \rangle =0$, $\langle 0|\theta|0\rangle =1$ and $\langle 1|\theta|1\rangle = 0$.  The operator $\pi_\theta$
can be represented as $\pi_\theta|0\rangle = 0$, $\pi_\theta|1\rangle = -i|0\rangle$. Then we  find $\pi_\theta^2 = 0$ and
$\{\pi_\theta,\theta\} =- i$. (Note that $\pi_\theta$ is antiHermitian.)
The state $|0\rangle$ satisfying $\pi_\theta|0\rangle =0$ indeed has zero norm: $\langle 0|0\rangle = 0$.

\section{The flat-torus case}\label{appendix:flat-torus}

In this appendix
we discuss some results of this paper in the special case where the spacetime corresponds to
a static flat torus.  We let the torus have volume $V$.   The normalized positive-frequency zero mode with small mass $M$ is
\bea
\varphi_{(0)}(t) & = & \frac{1}{\sqrt{2VM}}e^{-iMt} \nonumber \\
& \approx & \frac{1}{\sqrt{2VM}}\left( 1  - \frac{M^2}{2}t^2\right) - i \sqrt{\frac{M}{2V}} t.
\eea
Thus, the constants $B(M)$ and $C(M)$ in Eq.~\eqref{BMCM} are both $\sqrt{MV/2} = O(M^{1/2})$.

Next let us derive Eq.~\eqref{delta-result} for this
special case. The zero-mode contribution to the FP-ghost propagator with the IR divergences regularized with small mass $M$ is
\bea
D^{(0)ab}(t,t') & = & \frac{i}{2 M V}\delta^{ab}\left\{ \theta(t-t') \exp\left[ - iM(t-t')\right]\right. \nonumber \\
&& \,\,\,\,\,\, \left. + \theta(t'-t)\exp\left[ iM(t-t')\right]\right\} \nonumber \\
& = & \frac{i}{2MV}\delta^{ab}\exp\left( -iM|t-t'|\right)\nonumber \\
& \approx &  \delta^{ab}\left(\frac{i}{2 M V} + \frac{1}{2V}|t-t'|\right).
\eea
Then
\be
D_{\textrm{eff}}^{(0)ab} = \frac{1}{2V}\delta^{ab}|t-t'|.
\ee
Hence
\be
\frac{\partial^2\ }{\partial t\partial t'}D_{\textrm{eff}}^{(0)ab} =  - \frac{1}{V}\delta^{ab}\delta(t-t'),
\ee
which is Eq.~\eqref{delta-result}.

Next, we show that the requirement of time-translation invariance of the vacuum state for the theory with nonzero gauge parameter $\xi$ in the
Lagrangian density~\eqref{lagrangian1} leads to a divergent zero-mode two-point function~\cite{MiaoTsamisWoodard}.
This is in contrast to the de~Sitter case we study in the next Appendix.

Since the two-point function is diagonal in the Lie-algebra space, we study it for the Abelian theory.   For $\xi\neq 0$, the field equation with $q=0$ for the gauge
field $A_\mu$ on the static flat torus is
\be
\partial^\mu (\partial_\mu A_\nu - \partial_\nu A_\mu) + \frac{1}{\xi}\partial_\nu \partial_\mu A^\mu = 0.
\ee
With the assumption $A_i = 0$ and that $\partial_i A_0 = 0$, $i=1,2,\cdots,n-1$, we have $d^2 A_{0}/dt^2 = 0$, which can be solved as
\be
A_{0(0)}(t) = \frac{1}{\sqrt{V}}\left( \hat{Q} +  t\hat{P}\right),  \label{A0-mode}
\ee
where $\hat{Q}$ and $\hat{P}$ are constant operators.
 The equal-time commutation relation between $A_0$ and $\dot{A}_0$ reads
\be
\left[ A_0(t,\mathbf{x}), \dot{A}_0(t,\mathbf{x}')\right] = - i \xi \delta(\mathbf{x},\mathbf{x}').
\ee
By integrating over $\mathbf{x}$ and $\mathbf{x}'$ we find
\be
\left[ A_{0(0)}(t), \dot{A}_{0(0)}(t)\right] = - \frac{i\xi}{V}. \label{commuA0}
\ee
This implies $[\hat{Q},\hat{P}] = - i\xi$.
It is clear from Eq.~\eqref{A0-mode} that, if we require that the perturbative vacuum state $|0\rangle$ be time-translation invariant, then
$\hat{P}|0\rangle = 0$.  This means that $\langle 0|[A_{0(0)}(0)]^2|0\rangle = V^{-1}\langle 0|\hat{Q}^2|0\rangle = \infty$, which can be inferred
from Heisenberg's uncertainty relation $\langle 0|\hat{Q}^2|0\rangle\langle 0|\hat{P}^2|0\rangle \geq \xi^2/4$.  For the Landau gauge with $\xi=0$ the
field equations imply $\hat{P}=0$, and it is possible to require the condition $A_{0(0)}(t)|0\rangle = 0$ as described in Sec.~\ref{sec:conserved-charges}.

\section{The gauge-field zero-mode condition for the de~Sitter case}\label{appendix:zero-mode-for-A}

In this appendix we show that the condition $\varpi_{\tilde{B}(0)}|0\rangle=0$ is automatically satisfied by the perturbative Bunch-Davies
vacuum state $|0\rangle$ in the Landau gauge.  This means that there will be no need to remove the zero-mode contribution from the
gauge-field propagator by hand.   We omit the Lie-algebra indices in this appendix as well.

We first note that, under the assumption that $A_0$ is space-independent and that $A_i=0$, the free-field equation with $\xi\neq 0$,
\be
\nabla^\nu (\nabla_\nu A_\mu - \nabla_\mu A_\nu) + \frac{1}{\xi}\nabla_\mu \nabla_\nu A^\nu =0,
\ee
becomes
\be
\frac{1}{V(t)}\frac{d\ }{dt} [V(t)A_0(t)] = \textrm{const}.
\ee
(Notice that $\nabla_\nu A_\mu - \nabla_\mu A_\nu = 0$ under the assumptions we have made.)
Thus, the general solution is a linear combination of $[V(t)]^{-1}$ and $[V(t)]^{-1}\int_0^t dt'\, V(t')$.   It can be shown that the space-independent
positive-frequency solution, which is the coefficient of the annihilation operator,  is proportional to $\nabla_\mu F_0$, where $F_0$ is the positive-frequency
solution to the scalar field equation given by Eq.~\eqref{answerF}.
Writing this solution as $(A_0,A_i) = (\Phi(t),0)$ and provisionally normalizing $\Phi(t)$ by requiring
\be
i\int d\mbox{\boldmath$\theta$}\,\sqrt{|g|}[\Phi^*(t)\nabla^0 \Phi(t) - \Phi(t)\nabla^0 \Phi^*(t)]  = 1,
\ee
we find
\be
\Phi(t) = \frac{1}{\sqrt{2c_0}V(t)}\left[ \int_0^t dt'\,V(t') + i c_0\right].
\ee

Now, as in the previous Appendix the equal-time canonical commutation relations lead to
\be
\left [A_{0(0)}(t),\dot{A}_{0(0)}(t)\right] = - \frac{i\xi}{V(t)}.
\ee
By writing
\be
A_{0(0)}(t) = \Phi(t)a + \Phi^*(t) a^\dagger, \label{A0-Phi}
\ee
where the perturbative vacuum state $|0\rangle$ satisfies $a|0\rangle = 0$, we obtain
$[a,a^\dagger] = -\xi$.  By expressing the annihilation operator $a$ in terms of $A_{0(0)}(t)$ we can write the condition
$a|0\rangle = 0$ as
\be
\left\{ \Phi^*(t) \frac{d\ }{dt}\left[ V(t)A_{0(0)}(t)\right] - \frac{V(t)}{\sqrt{2c_0}}A_{0(0)}(t)\right\} |0\rangle = 0 \label{conditionA}
\ee
for all $\xi\neq 0$.
This condition is $\xi$-independent, and it is natural to require it for $\xi=0$ as well.  Now, recall that
$V(t)A_{0(0)}(t) = Q_A$ is a conserved Noether charge if $\xi=0$
[see Eq.~\eqref{Q_A-def}], which implies $d[V(t)A_{0(0)}(t)]/dt = 0$.  By using this operator equation in Eq.~\eqref{conditionA}
we find, recalling that $A_{0(0)} = - \varpi_{\tilde{B}(0)}$, that $\varpi_{\tilde{B}(0)}|0\rangle  =0$.
(We note in passing that the conservation of $Q_A$ implies $a^\dagger = -a$.)

Now, we readily find
\be
\langle 0|\left[A_{0(0)}(0)\right]^2|0\rangle = - \frac{\xi c_0}{2[V(0)]^2} \label{A00-xi}
\ee
from
Eq.~\eqref{A0-Phi} with
\be
\Phi(0) = i \sqrt{\frac{c_0}{2}}\,\frac{1}{V(0)}
\ee
and the commutation relation $[a,a^\dagger] = -\xi$.
In the rest of this appendix we show that Eq.~\eqref{A00-xi} results from the known
two-point functions for the massless vector field in the literature~\cite{AllenJacobson,TsamisWoodard2007,Youssef, FroebHiguchi} as a consistency check.

The two-point function given by Eq.~(45) of Ref.~\cite{FroebHiguchi} for $n\geq 4$
can be written as
\bea
\Delta_{\mu\nu'}(x,x') & = & \frac{n-2}{n-3} H^{-2} D_{M_0}(Z)\partial_\mu \partial_{\nu'}Z \nonumber \\
&& + \frac{1}{n-3}H^{-2}\left[\frac{d\ }{dZ}D_{M_0}(Z)\right](\partial_\mu Z)(\partial_{\nu'}Z) \nonumber \\
&&  + \left(\xi - \frac{n-1}{n-3}\right)\partial_\mu\partial_{\nu'} \tilde{D}(Z), \label{Deltax-x}
\eea
where $M_0 = \sqrt{n-2}H$.
Here, the variable $Z$ is defined by
\bea
Z  & = & \cos H\mu(x,x') \nonumber \\
& = & - \sinh Ht \sinh Ht' + \cosh Ht \cosh Ht' \cos \chi(x,x'), \nonumber \\
\eea
where $\mu(x,x')$ is the spacelike geodesic distance between the two points $x$ and $x'$
and where $\chi(x,x')$ is the angle on $S^{n-1}$ between the space components of
these two points. The function $D_M(Z)$ is the two-point function for the minimally coupled scalar field with mass $M$, and the function
$\tilde{D}(Z)$ is defined by
\be
\tilde{D}(Z) = - \lim_{M\to 0^+} \frac{\partial\ }{\partial M^2} \left[ D_M(Z)  - D_M(-1)\right].
\ee
We have $\partial_t Z|_{t=t'=0} =0$ and $\partial_t \partial_{t'}Z|_{t=t'=0} = -H^2$.

The normalized zero mode for
the minimally coupled massless scalar field with mass $M_0$ can be found from Eq.~\eqref{original} with $\ell=0$ and
$\lambda = (n-3)/2$ as
\bea
G_0(t)  & = &
H^{\frac{n-2}{2}}\left[\frac{\Gamma(n-2)}{2}\right]^{1/2}\frac{1}{H^{\frac{n-1}{2}}\sqrt{V(0)}}\nonumber \\
&& \times (\cosh Ht)^{-\frac{n-2}{2}}\mathrm{P}_{\frac{n-4}{2}}^{-\frac{n-2}{2}}(i\sinh Ht),
\eea
where $1/[H^{\frac{n-1}{2}}\sqrt{V(0)}]$ is the spherical harmonic on $S^{n-1}$ with $\ell=0$.  The zero-mode contribution to $D_{M_0}(Z)$ is
$G_0(t)G_0^*(t')$ whereas that to $D_M(Z)$ with small $M$ is $F_0(t)F_0^*(t')$, where $F_0(t)$ is given by Eq.~\eqref{answerF}.
Then
\bea
&& \langle 0|\left[A_{0(0)}(0)\right]^2|0\rangle\nonumber \\
&&  = - \frac{n-2}{n-3}|G_0(0)|^2\nonumber \\
&& \,\,\,\, - \left. \left( \xi - \frac{n-1}{n-3}\right)\lim_{M\to 0^+} \frac{\partial\ }{\partial M^2}|\partial_t F_0(t)|^2\right|_{t=0}.
\label{A00-expectation}
\eea
 Using 8.756.1 of Ref.~\cite{GradshteynRyzhik},
\be
\mathrm{P}_{\nu}^{-\mu}(0) = \frac{2^{-\mu}\sqrt{\pi}}{\Gamma(\frac{\nu+\mu}{2}+1)\Gamma(\frac{-\nu+\mu+1}{2})},
\ee
and using the doubling formula~\eqref{Gamma2x}, we find
\be
|G_0(0)|^2  = \frac{(n-1)c_0}{2(n-2)[V(0)]^2},  \label{G0value}
\ee
where the constant $c_0$ is given by Eq.~\eqref{c0def}.  We also find
\be
\left. \lim_{M\to 0^+}\frac{\partial\ }{\partial M^2}|\partial_t F_0(t)|^2\right|_{t=0} = \frac{c_0}{2[V(0)]^2}. \label{F0value}
\ee
By substituting Eqs.~\eqref{G0value} and \eqref{F0value} into
Eq.~\eqref{A00-expectation} we obtain Eq.~\eqref{A00-xi} as expected.  For $n=2$ and $3$
one can verify Eq.~\eqref{A00-xi} by explicit integration over the space using the two-point function found in Appendix C of Ref.~\cite{FroebHiguchi}.


\end{document}